\documentclass[12pt,preprint]{aastex}

\usepackage{xspace}
\usepackage{amsmath}
\usepackage{multirow}
\newcommand{\eight}{850~$\mu$m\xspace}
\newcommand{\four}{450~$\mu$m\xspace}
\newcommand{\clfind}{clfind2d\xspace}

\begin{document}

\title{High Mass Star Formation I: The Mass Distribution of Submillimeter 
Clumps in NGC~7538}

\author{Michael A. Reid and Christine D. Wilson}
\affil{Department of Physics and Astronomy, McMaster University, 
Hamilton, ON, L8S 4M1, Canada}

\begin{abstract} 

We present submillimeter continuum maps at \four and \eight of a
12\arcmin$\times$8\arcmin~region of the NGC~7538 high-mass star-forming
region, made using the Submillimeter Common-User Bolometer Array (SCUBA)
on the James Clerk Maxwell Telescope.  We used an automated clump-finding
algorithm to identify 67 clumps in the \four image and 77 in the \eight
image.  Contrary to previous studies, we find a positive correlation
between high spectral index, $\alpha$, and high submillimeter flux, with
the difference being accounted for by different treatments of the error
beam.  We interpret the higher spectral index at submillimeter peaks as a
reflection of elevated dust temperature, particularly when there is an
embedded infrared source, though it may also reflect changing dust
properties.  The clump mass-radius relationship is well-fit by a power law
of the form M~$\propto$~R$^{-x}$ with $x$ = 1.5--2.1, consistent with
theories of turbulently-supported clumps.  According to our most reliable
analysis, the high-mass end ($\sim$100--2700 M$_{\odot}$) of the
submillimeter clump mass function in NGC~7538 follows a Salpeter-like
power law with index $2.0 \pm 0.3$.  This result agrees well with similar
studies of lower-mass regions $\rho$ Oph and Orion B.  We interpret the
apparent invariance of the shape of the clump mass function over a broad
range of parent cloud masses as evidence for the self-similarity of the
physical processes which determine it.  This result is consistent with
models which suggest that turbulent fragmentation, acting at early times,
is sufficient to set the clump mass function.

\end{abstract}


\section{Introduction}

Although the form of the Galactic stellar initial mass function (IMF) has
been accurately determined in the decades since Salpeter's original paper
on the subject, a definitive explanation of its origin is still lacking
\citep{sal55,kroupa}.  Stars form from dense knots of dust and gas in
the clumpy interstellar medium (ISM).  Thus, to understand the origin of
the stellar IMF, we must first understand the structure of the clumpy ISM
and its origins.  The relatively recent advent of sensitive submillimeter
(submm) bolometer arrays has enabled us to begin studying the distribution
of cold, dense, dusty molecular clumps in star-forming regions.  
Extensive studies of the submm clump distribution in low-mass star forming
regions (e.g. \citealt{man98}, \citealt{dj2000b}, \citealt{m01})
demonstrate that the mass function of pre-stellar clumps very closely
resembles the Salpeter stellar IMF.  This similarity implies that the
stellar mass function is determined very early in the process of star
formation, at least at low stellar masses.  Turbulent fragmentation has
been suggested as one process which might be able to account for the
observed form of the clump mass function \citep{ef96,em97,kbb98,pn02}.

The structure of the cold, dusty ISM in regions of massive star formation
remains relatively unexplored, though its study bears on several important
problems in star formation.  For example, it would be very helpful to know
if the mass function of high-mass clumps is the same as that of low-mass
clumps.  Can turbulent fragmentation explain the origins of the clump mass
function even at very high masses, or must some other mechanism, such as
competitive accretion \citep{bon97}, be invoked?  Because they contain
clumps spanning several orders of magnitude in mass, massive star-forming
regions bridge the gap between the scales on which individual stars form
and the larger scales on which clusters form.  Moreover, such studies have
the potential to locate the cold, dense structures which may be the
high-mass equivalents of the low-mass ``Class 0'' cores.  Most previous
studies of the structure of the dusty component of the ISM have
concentrated on regions of low-mass, relatively non-turbulent star
formation, such as $\rho$ Oph (\citealt{dj2000b}, \citealt{man98}).  
Comparatively little is known about the clump mass function in the
turbulent, high-mass regime, partly due to observational challenges
primarily arising from the large average distances of high-mass
star-forming regions from the Sun.  In this and a subsequent paper (Papers
I and II), we use sensitive submm maps of two relatively nearby high-mass
star-forming regions to measure the mass function of their cold, dusty
clumps.  In this paper, we present a
12\arcmin$\times$8\arcmin~(9$\times$6.5~pc) map of the Galactic
star-forming region, NGC~7538, while Paper~II presents a similar study of
M17.

	NGC~7538 is one of the nearest and youngest massive star-forming
regions, whose proximity and relatively simple morphology make it an
attractive candidate for this study.  Located at a distance of 2.8 kpc
(\citealt{cgg78}, but see also \citealt{mc86}), NGC~7538 contains a
prominent \ion{H}{2} region centered on IRS~5 and surrounded by several
highly active star-forming complexes.  The brightest complex contains
several very luminous infrared sources, IRS 1-3, each of which supports a
young, compact \ion{H}{2} region (\citealt{ww74}, \citealt{rots81}).  
The IRS~9 and 11 regions show maser activity, while NGC~7538S, near IRS
11, contains a very young ($\lesssim 10^{4}$ yr) accreting ``Class 0''
high-mass protostar \citep{swf03}.  All three of these complexes show
significant CO outflow activity \citep{davis1998}.  NGC~7538 also contains
several relatively quiescent, clumpy filaments trailing away from the more
active centers, as seen in the NH$_{3}$ maps of \citet{z01}.  The apparent
youth and relative proximity of NGC~7538 make it an excellent candidate
for the study of the clump mass function in a high-mass star-forming
region.  Moreover, the quiescent regions at some distance from the
\ion{H}{2} region are promising locations for detecting the very cold,
dense clumps which may represent the earliest stages of massive star
formation.

	Several other groups have mapped the large-scale structure of
NGC~7538, in both spectral lines and continuum emission.  
\citet{kramer98} mapped an area of $\sim$21\arcmin$\times$24\arcmin~in
$^{13}$CO~(1$\rightarrow$0) and about half that area in C$^{18}$O
(1$\rightarrow$0); their data have a resolution of 50\arcsec~and are
sensitive to the total column of gas, but have a smaller dynamic range
than our continuum map.  \citet{z01} used the VLA to map a slightly
smaller region in NH$_{3}$ (1,1) and (2,2) with an
8\arcsec$\times$6\arcsec~beam.  However, because the \citet{z01} map is
interferometric, it is insensitive to structure on scales larger than
$\sim 1$\arcmin.  \citet{ss04} mapped an 8\arcmin$\times$8\arcmin~region
of NGC~7538 in \eight and \four continuum with SCUBA.  However, this early
SCUBA map was made by chopping along the scan direction with a single chop
throw, which can smear out structure along the scan direction.  Also, the
sensitivity of the map is relatively low (0.13 Jy beam$^{-1}$ at \eight
and 0.9 Jy beam$^{-1}$ at \four).  The increased sensitivity and greater
spatial coverage of our new SCUBA maps make them better suited to a study
of the large-scale distribution of cold, dense clumps of NGC~7538.

	Our analysis of the submillimeter clumps in the NGC~7538 region is
divided into three sections.  In section \ref{sec:obs}, we describe the
acquisition, calibration, and reduction of the data.  In section
\ref{sec:props}, we detail the methods used to determine the masses,
temperatures, sizes, and other properties of the clumps.  In section
\ref{sec:cmd}, we analyze and discuss the bulk properties of the 
clumps, including their mass-radius relationship and mass 
function.  We summarize and conclude in section \ref{sec:sum}.  

\section{Observations and Data Reduction}
\label{sec:obs}

The data were obtained on the nights of 2003 April 15, 16, 20, and June 16
using the Submillimeter Common-User Bolometer Array (SCUBA) at the James
Clerk Maxwell Telescope (JCMT).  The observed region is approximately
12\arcmin$\times$8\arcmin.  The map was made with SCUBA in scan-mapping
mode, using the standard chop throws of 30\arcsec, 44\arcsec, and
68\arcsec~in both RA and DEC.  The total on-source integration time was
approximately 7.5 hours. Pointing checks were performed once per hour and
sky dips once every hour or two.  The pointing accuracy varied from night
to night, between 1.5\arcsec~and 3\arcsec, but was typically about
2\arcsec.  The \eight and \four sky opacities were calculated using
polynomial fits to the combination of the JCMT skydips and the 225 GHz
zenith optical depth measurements from the Caltech Submillimeter
Observatory (CSO).  The atmosphere was stable and fairly dry on all four
nights, with mean 225 GHz zenith optical depths of 0.068, 0.063, 0.065,
and 0.066, respectively.  The mean residuals in the polynomial fits to the
CSO optical depth translate into a systematic uncertainty in the fluxes of
less than 1\%.

The data were reduced using SURF, the standard SCUBA data reduction
software \citep{surf}.  Map reconstruction from the six chop throws was
done using the ``Emerson2'' method \citep{emerson2,em2}, with pixels
2\arcsec~on a side.  The mean rms flux measured in emission-free regions
of the maps is 0.021~Jy~beam$^{-1}$ at \eight and 0.18~Jy~beam$^{-1}$ at
\four.  The half-power beam widths at \eight and \four were measured to be
15.3\arcsec~and~8.1\arcsec, respectively.  The maps were calibrated using
both primary (Uranus and Mars) and secondary (CRL2688) calibrators. Making
the conservative assumption that the night-to-night variations in the
gains calibration are entirely due to measurement error, we determine the
uncertainty due to the gains to be 12\% at \four and 4\% at \eight.

As discussed by \citet{dj2000a}, reconstruction of a SCUBA scan map image
from its component chop throws using the Emerson2 technique introduces
artifacts that should be removed prior to further analysis.  These
artifacts take the form of spurious structure in the image on scales
significantly larger than the largest chop throw (i.e. a few arcminutes,
in the present case).  Often, the presence of a very strong source, such
as NGC~7538~IRS1, will introduce regions of negative flux (so-called
``negative bowls'') in its vicinity.  Because our chosen clump-finding
algorithm, \clfind \citep{clfind}, functions best on images without such
negative bowls, we have taken steps to suppress them.  \citet{dj2000b}
showed that the simple addition of a constant flux to the map to eliminate
bowls does not adequately address all of the sources of spurious
long-wavelength structure in the image.  We employ a method similar to
that of \citet{dj2000b} to suppress the bowls, namely convolving each map
with a Gaussian twice the width of the largest chop throw and subtracting
the result from the original image.  Because the chop technique itself
screens out all emission on scales much larger than the largest chop
throw, this technique eliminates only the artificial long-wavelength
modes.  In order not to introduce new negative bowls in the image by
smearing out and then subtracting bright sources, we masked out all pixels
with $|S| >5\sigma$ before producing the smoothed image.  This technique
leaves the image rms unchanged and does not significantly change the
height of a given peak with respect to its local surroundings, but it
mostly removes the negative bowls.  This greatly facilitates the
process of clump-finding.  Comparisons between matching clumps in the
``flattened'' and unflattened images shows a difference in their 
integrated fluxes of only a few percent, well below the level of the 
calibration uncertainties.

We used the 6 cm continuum map of \citet{i77} to correct our data for 
radio continuum emission.  Using the Westerbork Synthesis Radio 
Telescope, \citet{i77} mapped the \ion{H}{2} region and sources IRS1-3 in 
NGC7538 at 5 GHz.  The synthesized beam of their map is $\sim 14$\arcsec, 
which nearly matches the 15\arcsec~resolution of our \eight map.  We 
derived free-free corrections for each clump by comparing its peak fluxes 
at \eight and 6 cm and assuming a scaling law of $S_{\nu} \propto 
\nu^{-0.1}$ for the radio continuum emission.  The worst contamination 
occurs in clump SMM34, which lies directly over the center of the optical 
\ion{H}{2} region.  We estimate the free-free contamination there to be 
$\sim 20$\%.  Similarly, we estimate the contamination of clump SMM48, 
which includes the compact \ion{H}{2} regions of IRS 1-3, to be $\sim 
8$\%.  Elsewhere, the free-free contamination is typically on the order of 
a few percent.  Free-free corrections were found to be negligible at 
\four.

Corrections for contamination of the \eight filter by CO(3$\rightarrow$2)  
line emission were derived according to the method of \cite{seaquist}.  
We used archival CO (3$\rightarrow$2) spectra of IRS~1 and IRS~9 to derive
corrections to the \eight fluxes of $\sim 3$\% and $\sim 10$\%, 
respectively.  The best available CO map of NGC~7538 is that of
\citet{davis1998}, who used the JCMT to make a $\sim$
7\arcmin$\times$5\arcmin~CO(2$\rightarrow$1) map of the central part of
NGC~7538 (roughly 40\% of the area of our map).  Under the assumption that
the CO(2$\rightarrow$1) traces the CO(3$\rightarrow$2) fairly closely, we
derived corrections to the \eight fluxes for the clumps covered by the
\citet{davis1998} map.  These are typically less than 15\%, and are likely
to be further reduced (by perhaps a factor of two or more) by the fact
that our observations are chopped, whereas the CO(2$\rightarrow$1) map
traces the total flux.  Because these corrections are
relatively small and only calculable for the roughly half of our clumps
covered by the \citet{davis1998} map, we have not applied any correction 
for CO contamination to the data.

	A final correction must be made to account for the contribution of
the error beam to the integrated flux of each clump, which is especially
significant at \four.  The JCMT error beam can vary significantly on
timescales as short as a few weeks, so we derived fits to the error beam
at \eight and \four using our own calibration observations of Uranus and
CRL~2688.  We produced azimuthally averaged intensity profiles for each
calibrator and then fit them with two Gaussians each: a narrow Gaussian of
high amplitude, corresponding to the primary beam, and a broad, low
amplitude Gaussian, corresponding to the error beam.  We attempted to fit
a third, very low amplitude Gaussian, but were unsuccessful due to
signal-to-noise limitations.  Our estimates indicate that this third
component contributes at most 4\% of the flux within a
120\arcsec~aperture, and is therefore negligible.  The results of our beam
fits are shown in Table \ref{tab:beams}.  The beam fits to Uranus and
CRL~2688 agree within the errors and are very similar to those reported by
\citet{hs00}.  Henceforth, we use the Uranus beam fits, which have smaller
uncertainties.  We correct the integrated flux of each clump for the
contribution of the error beam within an aperture of radius equal to
the clump's effective radius.  A clump's effective radius is defined as
the radius of a circle of equal area to the clump.  The maximum error beam
correction applied at \eight was 14\%, and the typical correction was
between 3 and 10\%.  Similarly, the largest correction applied at \four
was 19\% and the typical value was between 5 and 15\%.  The uncertainty in
the error beam correction itself was taken to be half the difference
between the corrections due to the maximal and minimal error beams, which
are defined by the uncertainty ranges in Table \ref{tab:beams}.

\section{Properties of the NGC~7538 Clumps}
\label{sec:props}

\subsection{The Filamentary Global Morphology of NGC~7538}

\begin{figure}
\begin{center}
\includegraphics[width=5.5in]{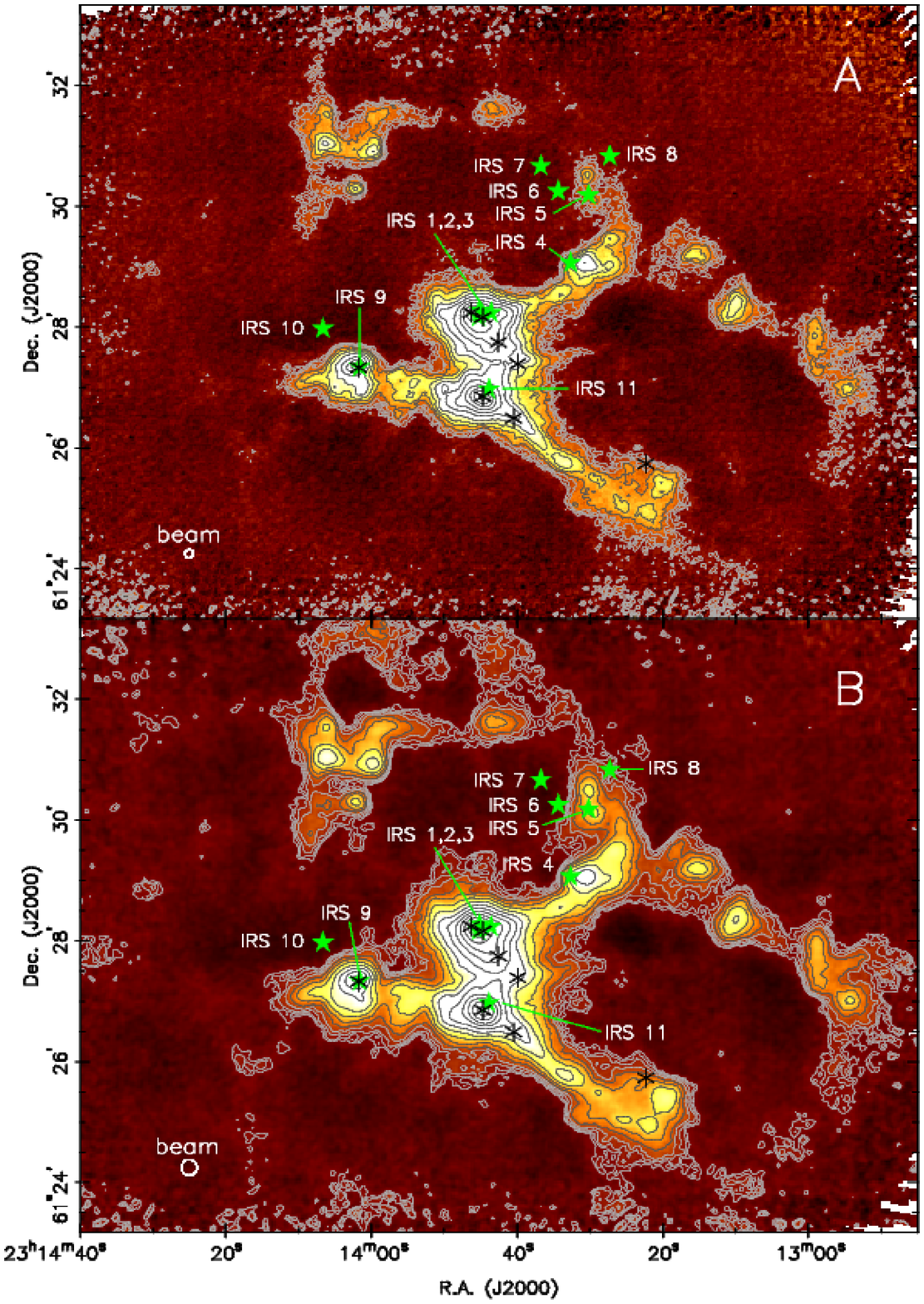}
\caption{Submillimeter continuum images of NGC~7538 at \four (A) and
\eight (B), in greyscale with logarithmically spaced contours.  The \four
contours begin at 3$\sigma$ (0.54 Jy beam$^{-1}$) and the \eight
contours begin at 3$\sigma$ (0.063 Jy beam$^{-1}$); both increase by
factors of 1.5.  Stars represent the positions of IRS point sources
(\citealt{ww74},\citealt{w79}). Asterisks represent the positions of water
masers from \citet{kam90}.  For visual clarity, only a representative
sample of the many masers in the IRS 1-3, 9, and 11 regions are shown.  
\label{fig:eightmap}}
\end{center}
\end{figure}

NGC~7538 is an excellent example of a filamentary star-forming region, as
can be seen in the maps of \citet{ss04} and, to a larger extent, in Figure
\ref{fig:eightmap}.  Most of the larger clumps are located along
filaments, although several smaller and apparently smooth filaments are
also visible.  The clumps associated with IRS~9 and 11 are connected in a
single long filament which stretches several parsecs southwest of IRS~11.
Another filament connects IRS~1-3, IRS~4, and IRS~5, then continues
southwest of IRS~5.  These two filaments may be joined in a bubble-like
structure by a faint bridge in the southwest corner of the image.

We see strong evidence of interaction between the optical \ion{H}{2}
region in NGC~7538, which is centered on IRS~5 (see Figure
\ref{fig:2mass}), and the submm-emitting material.  Although IRS~5 itself
is coincident with a submillimeter clump, most of the extended faint
emission from the \ion{H}{2} region sits in a submm void.  The steep
gradient in the submm walls of this void suggests compression by a wind
originating in the \ion{H}{2} region, which may have triggered much of the
present star formation activity in surrounding clumps.  

\subsection{Clump Identification Using \clfind}

In order that our method of identifying clumps be both systematic and
reproducible, we chose to use the popular \clfind program \citep{clfind}.
The \clfind algorithm typically works by contouring a map at three times
the image rms, $\sigma$, and at intervals of 2$\sigma$ above that.  We
found that, to improve the performance of the algorithm in this region
with considerable clustered substructure, it was necessary to set the
contour interval to $1\sigma$ when processing the \eight image.  Had we
not done so, many closely-spaced clumps would have been merged into
``plateaus'' by \clfind. We reprocessed the output of \clfind to merge
clumps whose peaks were separated by less than $2\sigma$, recovering a
similar (but not identical) result to that generated by the algorithm in
its default mode of operation. For the \four map, which has higher spatial
resolution, we found it sufficient to contour at the recommended $2\sigma$
level.  In this way, we identify very similar structures in both maps.

We emphasize that the choice of clump-finding algorithm introduces biases
into the clump properties.  For example, because \clfind attempts to
assign \emph{all} of the flux in an image to clumps, the masses and radii
of the some clumps are likely to be overestimated, particularly in highly 
clustered regions.  

\begin{figure}
\begin{center}
\includegraphics[width=5.5in]{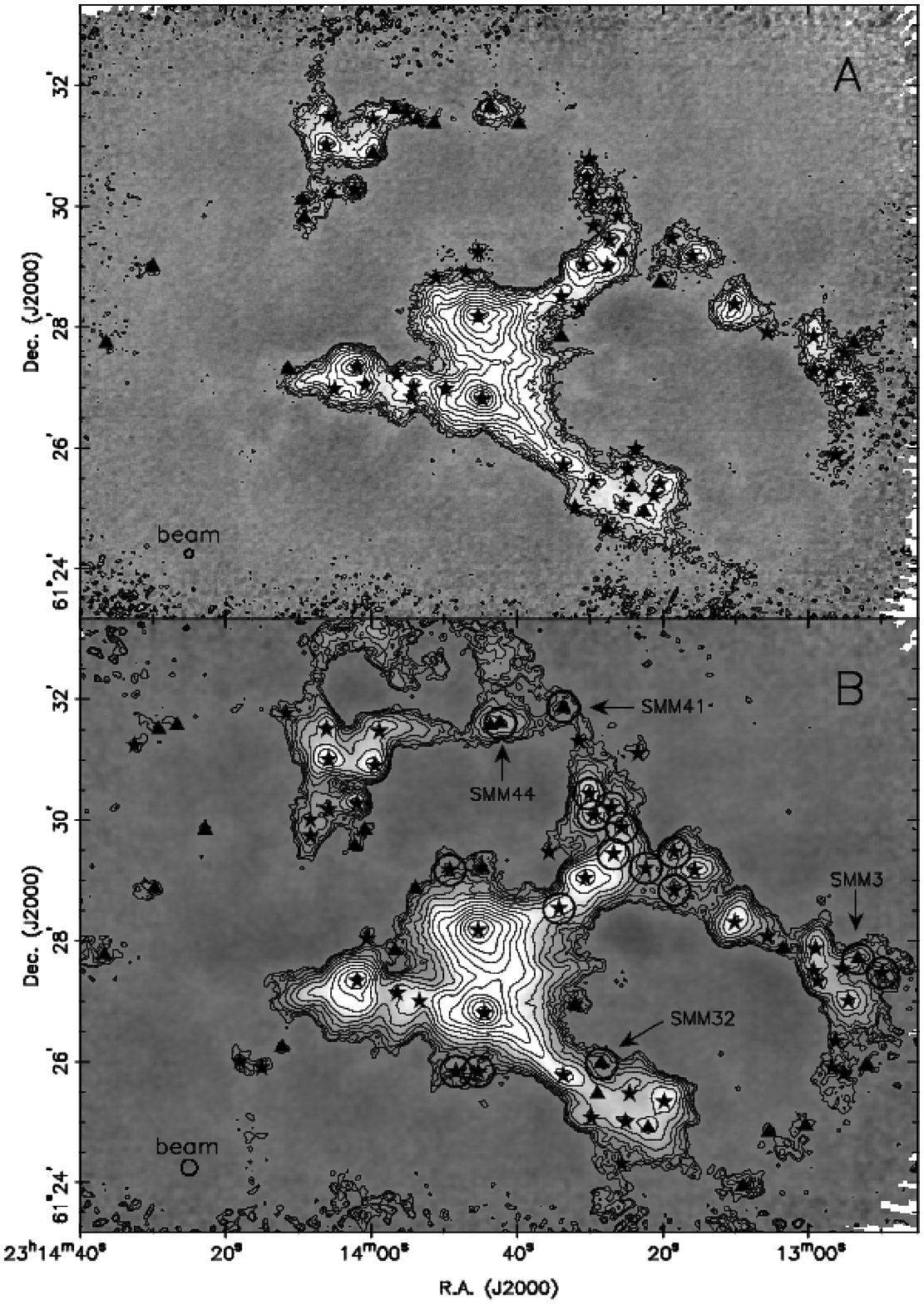}
\caption{The positions of the detected clumps in NGC~7538 at \four (A)
and \eight (B).  Contours are the same as in Figure \ref{fig:eightmap}.
Symbols indicate the peak positions of identified clumps which do
(stars) and do not (triangles) show coincidence with a 2MASS point source 
within their 0.5$S_{\rm peak}$ contour.  Circled symbols in the \eight image
indicate ``cold'' clumps (i.e. those with estimated dust temperatures, T$_{\rm 
dust} \le 30$~K).  Thus, a circled triangle indicates a 
possible cold, starless clump; the four such clumps are labelled. 
\label{fig:eightclumps}}
\end{center}
\end{figure}

In all, 77 clumps were identified in the \eight image and 67 in the \four
image.  These clumps and their derived properties are listed in Tables
\ref{tab:850clumps} and \ref{tab:450clumps}.  The positions of the clump
peaks are indicated in Figure \ref{fig:eightclumps}.  The differences in
clump numbers and boundaries between the two images arise because of the
higher signal-to-noise ratio (SNR)  of the \eight image and the higher
resolution of the \four image.  The angular resolution of the \four image
is nearly double that of the \eight image, so some of the \eight clumps
resolve into two or more clumps in the \four image.  For example, the
IRS~9 region, which appears as a single clump in the \eight image, is
resolved into four separate clumps in the \four image.  The high SNR of
the \eight image reveals several faint clumps not detected in the \four
image.  It also helps to distinguish closely spaced clumps with similar
peak fluxes, which may appear as a single elongated clump in the \four
image.

Comparison of the clump identifications from the \eight and \four images
reveals that 41 of the \eight clumps are spatially coincident with \four
clumps.  Of these 41 clumps, only 16 resolve into more than one clump in
the \four image, which has nearly double the resolution.  This comparison
suggests that most of the clumps seen in the \eight image are coherent
physical structures, rather than superpositions of unrelated smaller
structures.  Possible significant exceptions include the clumps
surrounding IRS 1-3 (SMM48) and IRS 9 (SMM60).  In both cases, the shapes
of the contours near the clump peaks suggest partially-resolved
substructure, but the flux differences among the sub-peaks are
sufficiently small that the \clfind algorithm groups them together.  
\citet{ss04} used the CLEAN algorithm to improve the resolution of their
continuum images of NGC~7538, achieving resolution of 7\arcsec~at \four.  
Their maps show more substructure in the IRS 1-3 and IRS 9 regions, and
may provide a clearer picture of the local source structure.  We interpret
their results cautiously, however, as their scan maps show evidence of
spurious elongation of structures along the scan direction.

\subsection{Clump Masses}
\label{sec:masses}

If the dust emission is optically thin, the total mass of a clump can be
calculated from

\begin{equation}
\label{eq:fluxtomass}
{\rm M}_{\rm clump} = 
\frac{S^{\rm int}_{\lambda}d^{2}}{\kappa_{\lambda}B_{\lambda}({\rm T}_{\rm dust})}~~, 
\end{equation}

\noindent where ${\rm M}_{\rm clump}$ is the mass of the clump, $S^{\rm
int}_{\lambda}$ is the flux of the clump at wavelength
$\lambda$ integrated over the boundary defined by \clfind,
$d$ is the distance to the clump, $\kappa_{\lambda}$ is the dust opacity
per unit mass column density at wavelength $\lambda$, and
$B_{\lambda}({\rm T}_{\rm dust})$ is the Planck function at wavelength
$\lambda$ and dust temperature T$_{\rm dust}$.

	Because determining both the temperature and the dust opacity for 
each clump individually is usually not possible, a number of simplifying 
assumptions must be made.  The simplest strategy is to assume that each 
clump can be represented by a single volume-averaged dust opacity and dust 
temperature, and that these values do not vary from clump to clump (e.g. 
\citealt{dj2000b}).  Another common strategy is to break clumps into 
categories and to assign a volume-averaged dust opacity and temperature to 
all of the clumps in each category (c.f. \citealt{man98}, 
\citealt{dj2000b}, \citealt{msl03}).  For this study, we assume that all 
clumps are represented by the same volume-averaged dust opacity.  By 
fitting spectral energy distributions, \citet{ss04} determined values of 
the dust emissivity index, $\beta$, of 1.2, 1.6, and 2 for the IRS 1, 11, 
and 9 clumps, respectively, although these values are considerably 
uncertain.  For simplicity, we will assume $\beta=1.5$ for all clumps.  
Hence, assuming a gas-to-dust ratio of 100 and taking the prescription 
from \citet{h83} that $\kappa_{\lambda} = 0.1 (250 \mu 
\mbox{m}/\lambda)^{\beta}$ cm$^{2}$~g$^{-1}$, we obtain 
$\kappa_{850\mu{\rm m}}=0.0087$~cm$^{2}$~g$^{-1}$ and 
$\kappa_{450}=0.031$~cm$^{2}$~g$^{-1}$.  This is the same prescription for 
the dust opacity used by \citet{ss04}.  Our values are also similar to the 
dust opacities used by \citet{m01}, who used $\kappa_{850\mu{\rm 
m}}=0.01$~cm$^{2}$~g$^{-1}$ for starless clumps and $\kappa_{850\mu{\rm 
m}}=0.02$~cm$^{2}$~g$^{-1}$ for protostellar envelopes in Orion, and by 
\citet{dj2000b}, who used $\kappa_{850\mu{\rm m}}=0.01$~cm$^{2}$~g$^{-1}$ 
for all of their detected clumps in $\rho$ Oph.

	The remaining free parameter in Equation \ref{eq:fluxtomass} is
the dust temperature.  Again, the simplest approach is to assume that all
clumps are isothermal and share the same temperature.  In crowded massive
star-forming regions, where each clump may contain one or more strong
central heating sources, the assumptions of an isothermal equation of
state and an invariant temperature are certainly approximate.  
Nonetheless, lacking suitable data for a more detailed analysis (and for
ease of comparison with previous work), we adopt this approach.  By
performing SED fits, \citet{ss04} found dust temperatures for the IRS 9
and IRS 11 regions of 35~K.  To facilitate comparison of our results with
theirs, we adopt 35~K as our universal dust temperature.  The masses thus
calculated are given in Tables \ref{tab:850clumps} (\eight clumps) and
\ref{tab:450clumps} (\four clumps).  The \eight masses range from 
1.4 to 2700 M$_{\odot}$ and the \four masses range from 4 to 3000 
M$_{\odot}$, straddling the regimes of stellar-mass objects and 
cluster-mass objects. 

	Our integrated fluxes and masses are considerably higher than
those of \citet{ss04}.  This likely due to a combination of the higher
sensitivity of our maps and to differences between the two studies in the
boundary definitions of highly clustered clumps.  Because \clfind attempts
to assign all of the flux in a map to clumps, the higher sensitivity of
our maps means that our integrated fluxes include more of the extended
emission from each clump.  In cases where there is a one-to-one
correspondence between the \eight and \four clumps and good agreement
between the \clfind-determined boundaries, the mean difference between the
two mass estimates is 12\%.  For example, the masses we derive for the
NGC~7538S/IRS11 clump (SMM46 and SMM46A, by our notation) are $2200 \pm
300$ M$_{\odot}$ at \eight and $2300 \pm 400$ M$_{\odot}$ at \four.  
Similarly, for clumps SMM63 and SMM57 (the two adjoining bright clumps in
the northeast corner of the image), we derive \eight masses of $140 \pm
20$ M$_{\odot}$ and $110 \pm 10$ M$_{\odot}$ respectively, compared to
\four masses of $140 \pm 20$ M$_{\odot}$ and $100 \pm 20$ M$_{\odot}$.  
Where the agreement between the clump boundaries is poor, the mean
difference in masses rises to 33\%.

\subsection{Dust Properties: Spectral Index and Temperature}
\label{sec:tvar}

	By dividing one of our maps by the other, we can produce a map of
the spectral index variations within NGC~7538.  The spectral index,
$\alpha$, takes the form,

\begin{equation}
\alpha = 
\frac{\log(S_{450\mu{\rm m}}/S_{850\mu{\rm m}})}{\log(850/450)}~~.
\label{eq:alpha}
\end{equation}

\begin{figure}
\includegraphics[width=\columnwidth]{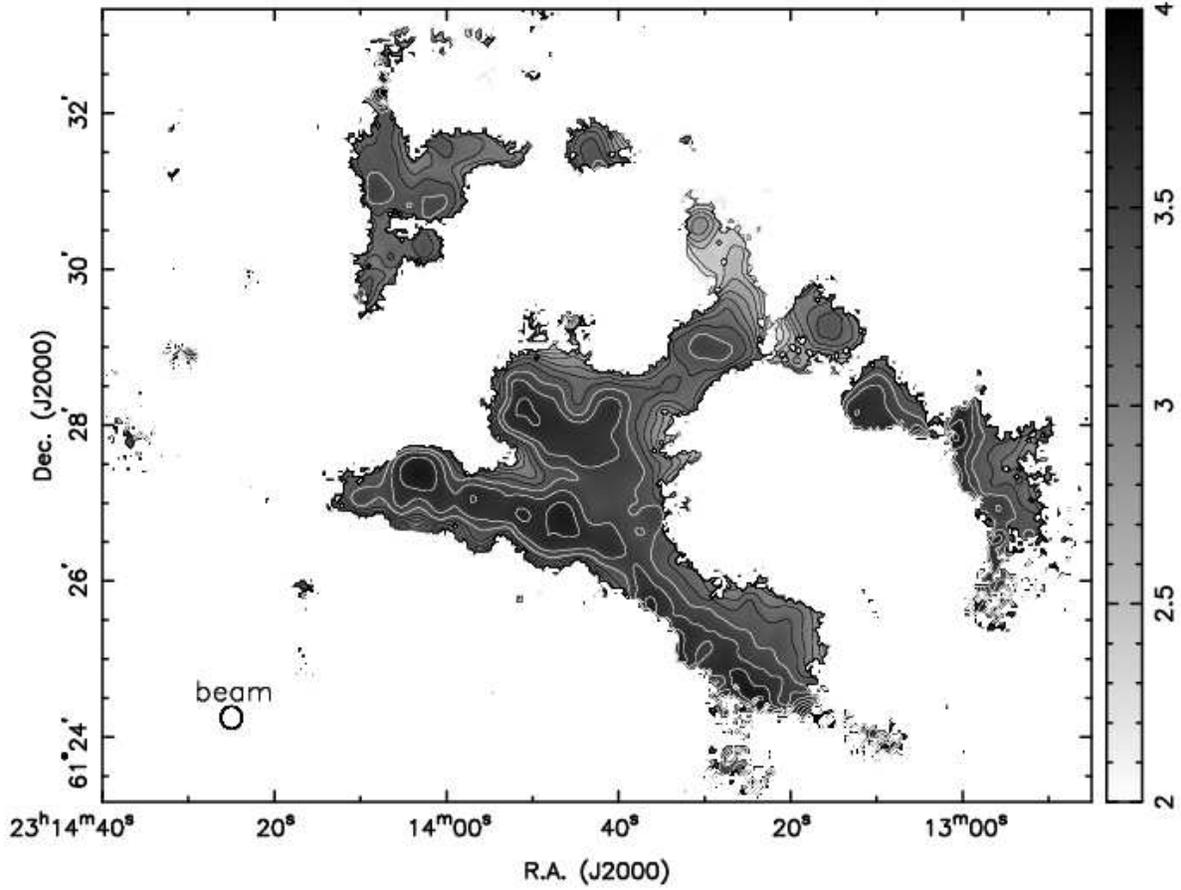}
\caption{Map of the spectral index in NGC~7538, produced by convolving the
\eight and \four images with each others' beams and dividing the two
images.  The effective beam size of the map is $\sim$17\arcsec.  The
contours are linearly spaced at intervals of $\alpha = 0.15$; the lowest
white contour is $\alpha = 3.4$.  Only pixels in which emission was
detected at $> 3\sigma$ in both filters are represented in the ratio map.
\label{fig:specmap}}
\end{figure}

\noindent Before dividing one image by another, we must ensure that they
have not only a common resolution, but a common beam structure.  Because
the error beams at \eight and \four differ so much (see Table
\ref{tab:beams}), it is not sufficient to simply match the main beam
structure by convolving the \four image to the same resolution as the
\eight image.  To produce \eight and \four images with fully matched
beams, we convolved each of the ``flattened'' images with the beam at the
opposing wavelength.  We then divided the convolved images and masked out
all pixels which were not detected at $\geq 3\sigma$ in \emph{both}
original images.  The result is a ``ratio map'' with spatial resolution of
$\sim$~17\arcsec (slightly coarser than the \eight image).  The
cross-convolutions alter the flux scaling in each map, so we must
re-calibrate the ratio map.  To derive the appropriate calibration factor,
we applied the same set of convolutions to the Uranus images and compared
the derived flux ratio with the one expected from tabulated fluxes for
Uranus at the time of our observations.  This gave a correction factor of
2.18, which we applied to the $S_{450\mu{\rm m}}/S_{850\mu{\rm m}}$ map.  
The flux ratio map was converted to a spectral index map using Equation
\ref{eq:alpha}; the result is shown in Figure \ref{fig:specmap}.

	We find a fairly consistent correlation between high submillimeter
flux and high spectral index, particularly in the bright regions around
IRS 1-3, 9, and 11.  Notable exceptions to this trend include the large
group of clumps in the southwest corner of Fig. \ref{fig:specmap} and
the group at the western edge of the image.  In both cases, there is a
fairly smooth gradient in $\alpha$ which does not reflect the locally
peaky nature of the submillimeter continuum emission (cf. Fig.  
\ref{fig:eightmap}).  Both regions border the bubble-like structure which 
dominates the western half of the region.
	
	Note that this correlation between high $\alpha$ and high
continuum flux is the \emph{opposite} trend to that found by \citet{fms01}
in their SCUBA study of star-forming region NGC~7129.  We believe the
discrepancy can be explained by the different methods used to account for
the error beam.  \citet{fms01} account for the error beam by adding a
constant level of 5\% of the peak flux to the \four map before convolving
it to the same resolution as their \eight map and dividing.  If we follow
the same procedure with our data, we also find a consistent
\emph{anti-}correlation between submillimeter continuum flux and spectral
index.  However, this technique for constructing an $\alpha$ map does not
account for the spatially-varying contribution of the \four error beam,
and hence results in comparison between images with mis-matched beams.

	For a given pair of frequencies, the spectral index, $\alpha$,
depends on the dust emissivity, $\beta$, and a temperature-dependent
correction to the Rayleigh-Jeans form of the Planck function (see Eqs. 5
and 6 of \citealt{fms01}).  An increase in $\alpha$ may arise due to an
increase in either or both of T and $\beta$.  For those clumps known to
have embedded infrared point sources, such as IRS 1-3, 9, and 11, the
observed increase in $\alpha$ toward the submillimeter peak may plausibly
be ascribed to an increase in the dust temperature.  Larger-scale
gradients, such as those described above, may also reflect temperature
gradients.  Alternatively, they could be the result of changing dust
properties (the formation of larger or smaller grains, for example).

	In section \ref{sec:masses}, we assumed a spatially invariant dust
emissivity index $\beta=1.5$.  If we again make this simplifying
assumption, we can use the ratio map to calculate the temperature of any
submillimeter-emitting parcel of dust via \citep{mitchell01}:

\begin{equation}
\label{eq:ratio}
\frac{S_{450_{\mu{\rm m}}}}{S_{850_{\mu{\rm m}}}} = \frac{e^{16.9/T} 
-1}{e^{32.0/T} -1}\left(\frac{850}{450}\right)^{3+\beta}~~.
\end{equation}

For those \eight clumps with five or more pixels in the ratio map, we used
Equation~\ref{eq:ratio} to convert the mean flux ratio in these pixels to
a mean dust temperature for the clump.  (Due to the significant resolution
mismatch between the \four map and the ratio map, this technique is not
applicable to the \four clumps.)  Equation \ref{eq:ratio} asymptotes to a
maximum flux ratio for a given $\beta$ (approximately 9.3 for $\beta =
1.5$).  No temperatures were calculated for clumps with mean flux ratios
above this value (the existence of such high flux ratios is likely a
reflection of a spatially varying $\beta$).  Subject to these constraints,
we are able to estimate mean temperatures for 42 of the 77 original \eight
clumps.  These temperatures appear in Table \ref{tab:850clumps}.  The
range of estimated temperatures is 8--329~K, with a median of 35~K.
Typical uncertainties on these temperatures, derived from the systematic
errors in the gains, sky opacities, and error beam fits, are $\pm$5~K for
T~$\lesssim~20$ K and $^{+20}_{-10}$~K for 20~K$<$T$<$ 50~K.  Above 50~K,
all we can say with confidence is that the clump is hotter than
$\sim$30~K.  In this sense, our temperature estimates can be interpreted
as lower limits on the real temperature of each clump. Clump masses
derived using these estimated temperatures and Eq.  \ref{eq:fluxtomass}
are also given in Table \ref{tab:850clumps}.

\subsection{Evolutionary States: Correlations with Signposts of Massive 
Star Formation}
\label{sec:stars}

	The NGC~7538 region is known to contain several objects in the
earliest stages of massive star formation.  \citet{swf03} give strong
evidence from both continuum and line observations for the presence of a
candidate ``Class 0'' massive protostar in the NGC~7538S region (near IRS 
11).  
The object contains a Keplerian disk approximately 30,000 AU in extent
with mass $\gtrsim 100${\rm M}$_{\odot}$, surrounding a central object of
mass $\sim40${\rm M}$_{\odot}$.  This object powers a strong outflow whose
dynamical age is estimated to be less than 10$^{4}$ yr.  Similarly, CS
(2$\rightarrow$1) observations by \citet{davis1998} reveal powerful
outflows emanating from the IRS 1 and 9 regions, with dynamical ages of
1.5$\times 10^{4}$ yr and 4.2$\times 10^{4}$ yr, respectively.  This
evidence strongly suggests that NGC~7538 is a site of vigorous, ongoing
massive star formation and motivates us to probe the evolutionary states
of its cold, dusty clumps.

	Maser surveys offer one tool for locating potential sites of massive
star formation.  In Figure \ref{fig:eightmap}, we have plotted the positions
of 22~GHz H$_{2}$O masers detected by \citet{kam90} in their survey of the
central 8\arcmin$\times$8\arcmin~of NGC~7538.  These H$_{2}$O masers trace all
of the known sites of maser emission in NGC~7538 (including those in other
transitions, such as methanol and OH), most of which are concentrated in the
IRS~1-3, 9, and 11 regions.  The notable exception is the maser which lies
roughly 3\arcmin~southwest of IRS~11, which is the brightest H$_{2}$O maser in
the region.  It lies within the \eight clump SMM26 and the \four clump SMM26B,
though not at the peak of either.  It may represent the location of an as-yet
unidentified massive protostar.  All of the other known sites of maser
emission fall within the boundaries of the submillimeter clumps associated
with IRS~1-3, 9, and 11.

\begin{figure}
\includegraphics[width=\columnwidth]{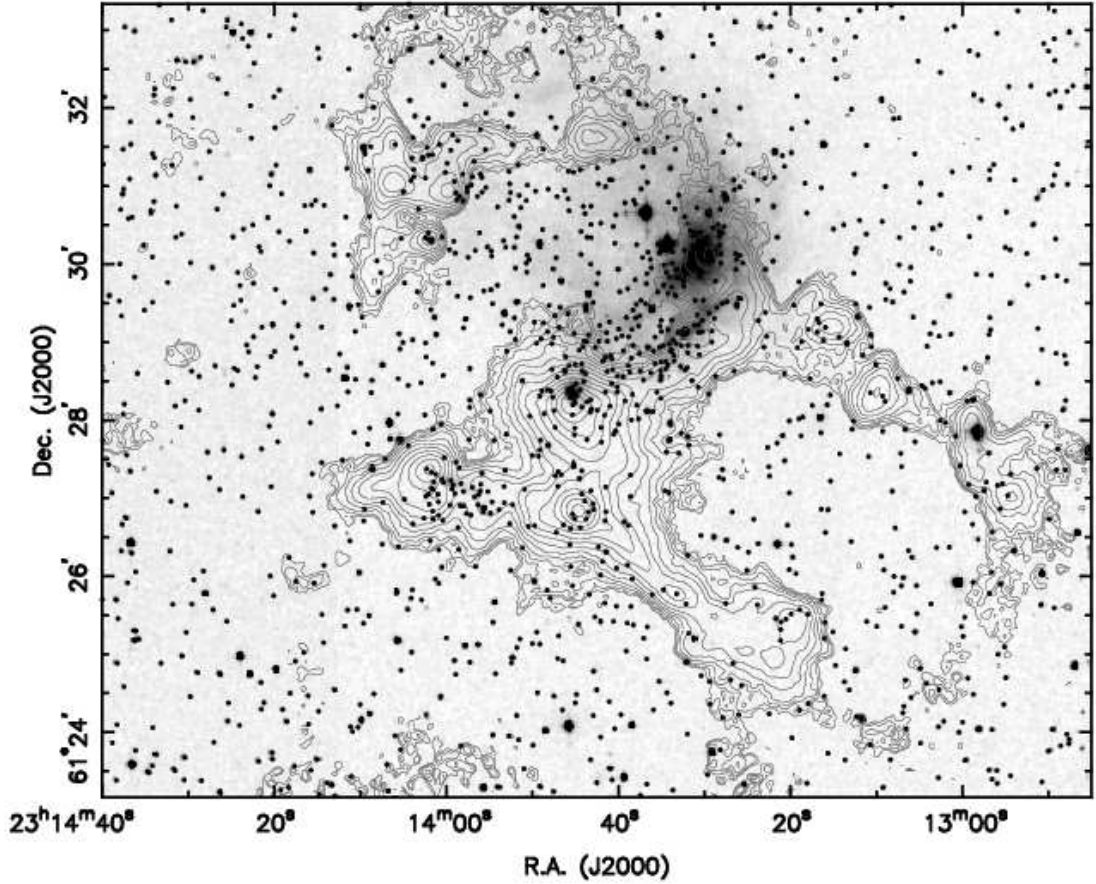}
\caption{Comparison of \eight submm emission (contours, as in Fig.  
\ref{fig:eightmap}) with 2MASS J image (greyscale) and 2MASS point sources
(black dots).  The \ion{H}{2} region is the diffuse feature filling the submm
void to the north of the prominent IRS 1-3 group.  Note the significant
over-density of infrared point sources in the IRS 1-3, 9, and 11 regions.  
Of the \eight clumps, 69\% show coincidence of a point source
within their 0.5$S_{\rm peak}$ contour. \label{fig:2mass}} 
\end{figure}

	We used the 2MASS maps of NGC~7538 (Fig. \ref{fig:2mass}) to
search for embedded infrared point sources, whose presence in a given
clump would indicate past or ongoing star formation.  Unfortunately, the
absence of such point sources cannot be taken as counter-evidence of star
formation; recent results from Spitzer show that, even for low-mass cores,
some ``starless'' cores actually contain heavily-extincted infrared
sources \citep{young04}.  The highest-resolution infrared data available
for the entire NGC~7538 region come from 2MASS (see Fig. \ref{fig:2mass}).  
The 2MASS exposures are deep enough to locate some of the embedded
sources, but are insufficiently deep to rule out definitively the
presence of an embedded infrared source in any particular clump.  In
Tables \ref{tab:850clumps} and \ref{tab:450clumps}, we specify the number
of 2MASS point sources which lie within the 0.5S$_{\rm peak}$ contour of
each clump.  Associations between submillimeter clumps and IRS sources
\citep{ww74} are listed in Table \ref{tab:assoc}.  Overall, 69\% of the
\eight clumps contain one or more such point sources, compared to 32\% of
the \four clumps;  however, this result does not necessarily mean that
69\% of the clumps have embedded protostars.  The density of 2MASS point
sources within the local field is about 60\% of the density of such
sources within the $> 3\sigma$ submillimeter-emitting region, so many of
the point sources coincident with submm clumps are certainly field stars.  
The exact fraction of ``starless'' clumps in our dataset is impossible to
determine with the data at hand but, based on the above analysis, we
estimate it to be around 50\%.

	We expect that the very youngest pre-stellar clumps will be those 
which are cold and which do not possess an embedded infrared point source.  
Examination of Table \ref{tab:850clumps} reveals several objects which are 
cold (mean T$\leq30$K) and contain no 2MASS point source within their 
0.5$S_{\rm peak}$ contour (clumps SMM3, 32, 41, and 44). These 
would be logical candidates for infrared and high-resolution spectral line 
and continuum follow-up.  There may be other cold, potentially starless 
clumps within the sample for which we were not able to calculate a 
temperature.

\section{Clump Mass Distributions: Evidence for Turbulent Fragmentation}
\label{sec:cmd}

\subsection{Clump Mass-Radius Relation}

\begin{figure}
\begin{center}
\includegraphics[width=4.5in]{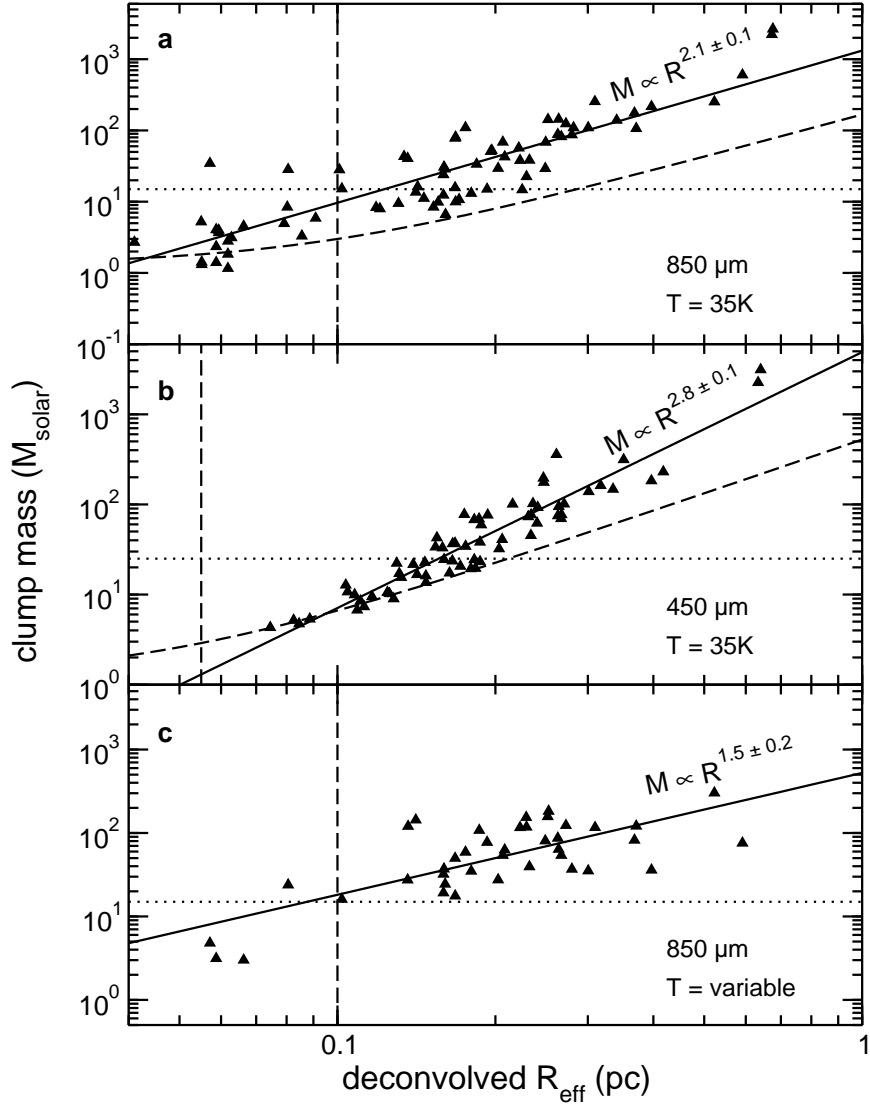}
\end{center}
\caption{Mass vs. deconvolved radius for three ensembles of clumps, with
masses calculated using the temperatures as indicated.  Panels (a), (b),
and (c) include 77, 67, and 42 clumps, respectively.  Horizontal dashed
lines indicate the estimated completeness thresholds of 15~M$_{\odot}$ at
\eight and 25~M$_{\odot}$ at \four.  The curving dashed lines in panels
(a) and (b) indicate the mass detection threshold as a function of
deconvolved radius.  Vertical dashed lines represent the beam radius;
clumps to the left of this line are unresolved.  Mass error bars are
omitted for clarity, but are typically close to the size of the plotted
points.  The solid lines are least-squares fits to power laws of the form
${\rm M}\propto R^{x}$.\label{fig:mvsr}}
\end{figure}

	Assessing the completeness of a sample of clumps extracted from a
submillimeter map is complicated by the fact that the clumps are not point
sources.  Hence, the detection limit is not an absolute cutoff in flux,
but rather a limiting rms surface brightness.  For example, a
spatially-extended high-mass clump, even with a large integrated flux, can
be more difficult to detect than a compact, low-mass object.  Therefore,
the mass detection threshold must be expressed as function of the
effective clump radius.  Figure \ref{fig:mvsr} shows the mass-radius
relation for the \eight and \four clumps with constant-temperature masses,
illustrating the dependence of the detection threshold on clump size.  We
define our smallest detectable clump as one with an area equal to at least
half that of the beam, whose flux is everywhere $\geq 3\sigma$ and whose
peak flux is $\geq 5\sigma$.  As seen in Fig. \ref{fig:mvsr}, it is
difficult to specify the mass above which the clump sample is complete.  
We estimate a lower completeness threshold of 15 M$_{\odot}$ at \eight and
25 M$_{\odot}$ at \four, although we emphasize the considerable
uncertainty in these estimates.

	Least-squares fits of the mass-radius data to a power law of the
form ${\rm M}\propto {\rm R}^{x}$ are shown in Figure \ref{fig:mvsr}.  
The \eight and \four constant-temperature mass-radius relations are best
fit by $x = 2.1 \pm 0.1$ and $2.8 \pm 0.1$, respectively.  A fit to the 42
clumps whose masses were calculated using the temperatures estimated from
the ratio map gives $x = 1.5 \pm 0.2$ (Fig. \ref{fig:mvsr}c).  None of the 
fits change substantially if we exclude the clumps below the estimated 
completeness threshold.

	Previous studies of the mass-radius relationship of submillimeter
clumps have concentrated on regions dominated by low-mass clumps, such as
NGC2068/2071 and $\rho$~Oph.  They have typically found $x\simeq 1$
\citep{m01}.  This is the expected result for an ensemble of
thermally-supported Bonnor-Ebert spheres \citep{b56,eb55}.  Similar
studies of the mass-radius relationship of CO clumps, spanning the mass
range $\sim 10^{-2}-10^3$ M$_{\odot}$ have typically found $x\simeq 2$
(\citealt{larson}, \citealt{ef96}).  Note, however, that this result only
emerges when the CO clumps from many regions are analyzed as an ensemble
\citep{ef96}; the power laws measured in each individual region have $x =
2.3-3.7$, with values $> 3$ being typical.  A value of $x=2$ is consistent
with the theoretical picture in which a turbulent molecular cloud
fragments, producing a fractal distribution of clump masses and sizes
(\citealt{ef96}, \citealt{em97}).  It is also the result predicted for an
ensemble of turbulently supported clumps, such as the logotropic spheres
of \citet{mp96}.

	We cannot rule out the possibility that the steep power law fit to
the mass-radius relation of the \four clumps in NGC~7538 may be
significantly affected by incompleteness at the low-mass end, where the
plotted data clearly intersect the detection threshold.  With no way to
confidently estimate the distribution of the points below the detection
threshold, we exclude the \four mass-radius relation from further
consideration.  The mass-radius relation for the \eight clumps with
constant-temperature masses is consistent with $x = 2$, which locates
these clumps in the turbulently supported regime, as expected for more
massive clumps.  As seen in Fig.~\ref{fig:mvsr}c, the mass-radius
relationship changes somewhat if we consider only those clumps for which
we are able to derive mean dust temperatures (i.e. those clumps with five
or more $\geq 3 \sigma$ pixels in both the \eight and \four maps, and with
relatively low mean spectral index).  The best-fit slope of $x = 1.5 \pm
0.2$ in this case is steeper than would be expected for a population of
Bonnor-Ebert spheres, but may be consistent with the interpretation of
these cores as dense, pre-stellar objects with significant turbulent
support.  This issue will be examined further in a forthcoming paper,
where we present high-resolution spectral line data.  The uncertainties
indicate that the differences between the fits in Figs.~\ref{fig:mvsr}a
and \ref{fig:mvsr}c are significant at the 2$\sigma$ level.  The primary
difference between the two data sets is that one attempts to account for
variations in temperature (via the spectral index).  This difference
demonstrates that the assumption of a universal dust temperature can have
a determining influence on conclusions about basic clump properties.

\subsection{Clump Mass Function}
\label{sec:cmf}

\begin{figure}
\begin{center}
\includegraphics[width=5.5in]{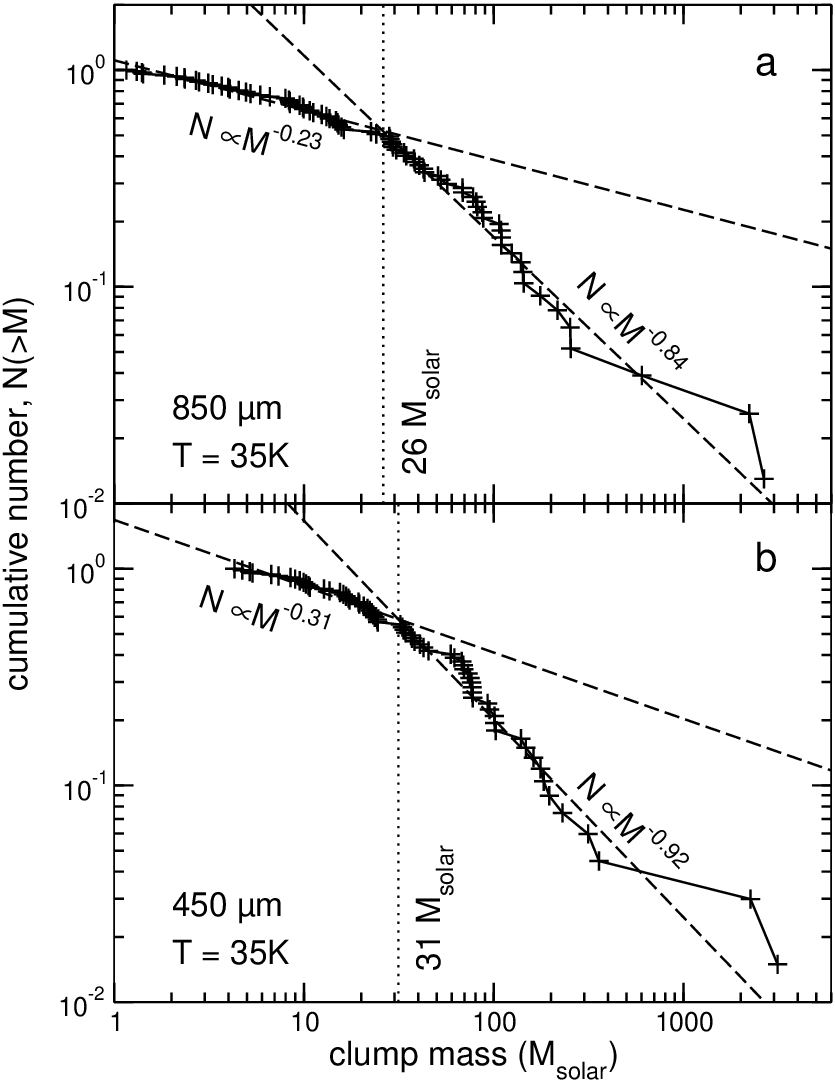}
\end{center}
\caption{Cumulative mass functions for (a) the 77 clumps detected at
\eight and (b) the 67 clumps detected at \four (b), with masses calculated
assuming T$=$35~K.  The dashed lines represent fits to broken power laws,
with slopes as labeled.  The vertical dotted lines indicate the break
points between the power laws.\label{fig:cumumf}}
\end{figure}

	To analyze the distribution of clump masses, we plot the
cumulative number of clumps with masses greater than M, $N(>{\rm M})$,
versus M.  Figure \ref{fig:cumumf} shows this cumulative mass function
(CMF) for the 77 clumps detected at \eight and the 67 clumps detected at
\four, with masses calculated assuming a universal T$_{\rm dust} = 35$~K.  
We have not computed a mass function for the \eight clumps with
variable-temperature masses due to concern that the omission of some
clumps may introduce bias into the mass function, particularly if there is
an intrinsic mass-temperature correlation not evident in our data.  
Omitting masses would be especially problematic in the case of the CMF,
whose overall shape depends on the position of each individual clump.  We
fit each CMF with a broken power law, which takes the form $N(M) \propto
M^{-\alpha}$ on either side of a break point (note that this $\alpha$ is
not the same as the spectral index discussed earlier).  The break point is
also a parameter of the fit.

	Both CMFs are well fit at the low-mass end by $\alpha
\sim$~0.2--0.3.  At the high-mass end, where incompleteness should be less
of a problem, both are well fit by $\alpha \sim$~0.8--0.9.  In both cases,
the break point occurs above our estimated incompleteness thresholds of
15~M$_{\odot}$ and 25~M$_{\odot}$ in the \eight and \four images,
respectively, although we again emphasize the uncertainty in the positions
of these thresholds.  Both the degree of flattening at the low-mass end of
the CMF and the position of the break point could be strongly influenced
by incompleteness.  Nonetheless, the existence of a break point above our
estimated incompleteness thresholds suggests that the change in slope is a
real property of the mass function of NGC~7538, not merely an artifact of
incompleteness.

	Both sections of the \four CMF (Fig. \ref{fig:cumumf}b) are
slightly steeper than their \eight counterparts (Fig. \ref{fig:cumumf}a).  
This may be a consequence of the fact that some of the \eight clumps
resolve into two or more smaller, lower-mass clumps at \four, shifting the
distribution toward lower masses.  The lower signal to noise ratio of the
\four map may also mean that it traces a smaller fraction of the mass of
each clump, again biasing the mass function to lower masses.  However, the
flux calibrations are considerably less certain at \four than at \eight,
so it is difficult to be certain whether the difference in slopes between
the \four and \eight CMFs is real.

\begin{figure}
\begin{center}
\includegraphics[width=5.5in]{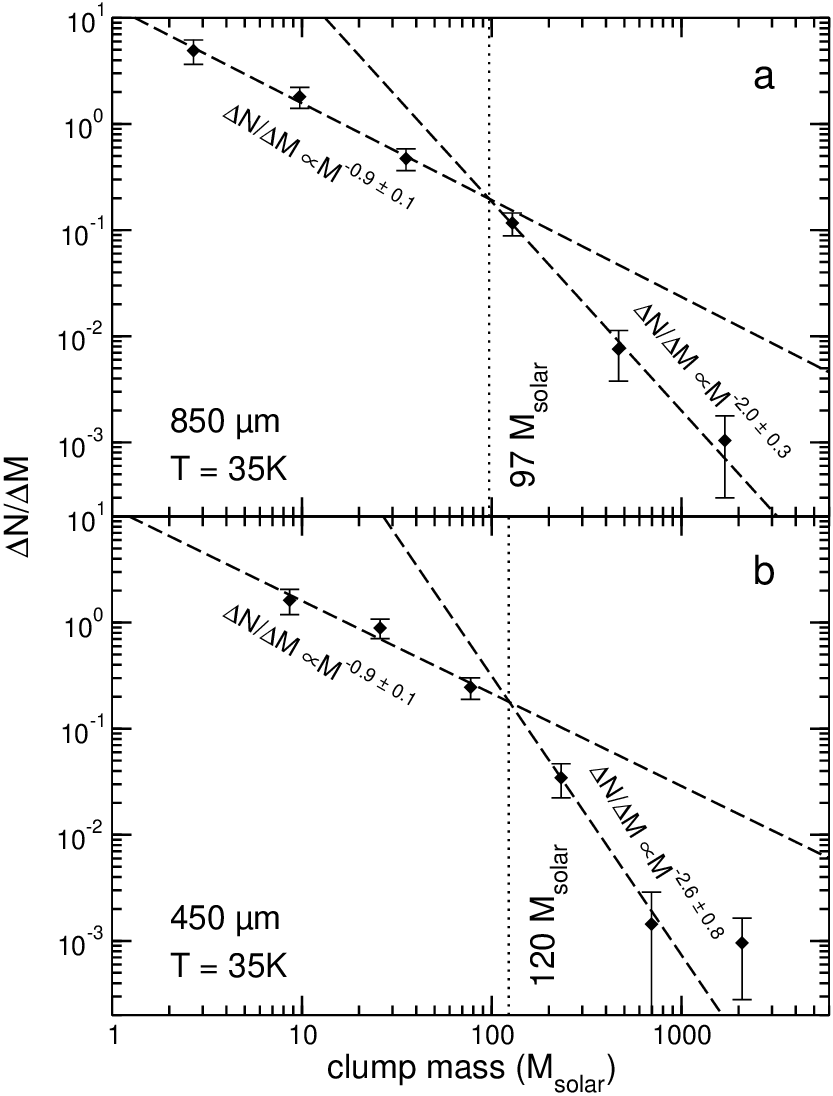}
\end{center}
\caption{Differential mass functions for (a) the 77 clumps detected at
\eight and (b) the 67 clumps detected at \four, with masses calculated
assuming T$=35$~K.  The dashed lines represent fits to broken power laws,
with slopes as labeled.  The vertical dotted lines indicate the break
points between the power laws.  \label{fig:diffmf}}
\end{figure}

	A possible criticism of the cumulative mass function is that its
shape is dependent on the position of each clump; there is no built-in
averaging-out of uncertainties.  With 77 clumps detected at \eight, we
have enough objects to construct a differential mass function, $\Delta
{\rm N}/\Delta {\rm M}$, where $\Delta $N is the number of clumps in the
mass bin $\Delta {\rm M}$.  Figure \ref{fig:diffmf} plots such a function
for the same sets of clumps as in Fig. \ref{fig:cumumf}.  Again, we fit to
two power laws of the form $\Delta {\rm N}/\Delta {\rm M} \propto {\rm
M}^{-\gamma}$ on either side of a fitted break point, where $\gamma =
\alpha + 1$.  At the low-mass end, we find $\gamma = 0.9 \pm 0.1$ for the
clumps at both wavelengths.  At the high-mass end, we find $\gamma = 2.0
\pm 0.3$ for the \eight clumps and $\gamma = 2.6 \pm 0.8$ for the \four
clumps.  Again, the position of the break point should be viewed as
considerably uncertain, particularly given how much higher it is for the
differential mass functions than for the cumulative ones.
 Comparing the differential and cumulative mass functions of Figs.
\ref{fig:cumumf} and \ref{fig:diffmf}, we see that they all agree within
the uncertainties at the high-mass end, where the data are most likely to
be complete.  

	In all cases, lower incompleteness at higher clump masses
means that the fit to the high-mass end of each mass function is more
reliable than that to the low-mass end.  The higher sensitivity and better
calibration of the \eight data make the \eight clump mass estimates more
reliable than the \four ones.  Finally, the straightforward interpretation
of the uncertainties in the differential mass function leads us to trust
it more than the cumulative mass function.  Thus, our most reliable
estimate of the power law exponent of the clump mass function in NGC~7538
is $\alpha = 2.0 \pm 0.3$, from the high-mass end of the \eight
differential mass function.

\subsection{Mass Functions and Turbulent Fragmentation}

	In Table \ref{tab:powerlaws}, we compare our results for NGC~7538
to similar mm and submm continuum studies of star-forming regions in which
the mass function is fitted with a broken power law.  All of these studies
should be sensitive to the cold, dense dust and gas that forms the
precursor structures to stars.  The method of extracting the clumps and
converting their integrated fluxes to masses is similar in all of the
studies.  Although both of the regions previously studied in this way
($\rho$ Oph and Orion B) have considerably lower average clump masses than
NGC~7538, the shape of the mass function in all three regions is
remarkably similar.  In particular, the high-mass end of the mass function
is well fit by an approximately Salpeter mass function ($\gamma = 2.35$,
\citealt{sal55}).  This agreement in the power-law exponents among the
various clump mass functions suggests that the clump mass function is
shaped by a common set of physical processes which operate self-similarly
across a broad range of clump and parent cloud masses.  This appears to be
true despite the fact that the star-forming future of any given clump is
usually not clear.  In the low-mass regions, many of the clumps may form
single stars, while others may form binaries and higher-order multiples.  
The NGC~7538 clumps span a range of masses, from those which may form only
a single star, to those such as the IRS 1-3, 9, and 11 clumps, which will
probably form small clusters of stars.  Nevertheless, it appears that the
Salpeter-like clump mass function may extend all the way from
stellar-scale to cluster-scale submm clumps.

	Turbulent theories of molecular cloud evolution suggest the
mechanism by which this situation may come about.  Hydrodynamic
simulations of turbulent, self-gravitating gas have shown that, for a
range of initial conditions corresponding to those typical of molecular
clouds, the shape of the mass function is independent of the total mass of
the cloud \citep{tilley}.  Taken together, the results of Table
\ref{tab:powerlaws} imply a shape for the clump mass function which does
not vary substantially from cloud to cloud, and whose intrinsic scales
(such as the position of the turnover) may be set by global properties of
the parent molecular cloud, such as the total number of Jeans masses
present.

	In Table \ref{tab:powerlaws}, we also list those studies of the
dust continuum clump mass function in Galactic star-forming regions in
which the mass function was fitted with a single power law.  In the case
of Serpens, which is a nearby low-mass star forming region probed at small
linear scales, the slope of the clump mass function is consistent with
that of Salpeter.  The remaining three regions (the Lagoon Nebula, KR 140,
and RCW 106) are all more distant high-mass star-forming regions which
were probed at spatial scales similar to or larger than those in our study
of NGC~7538.  The mass function in these regions is systematically more
shallow than both the Salpeter mass function and those found in $\rho$
Oph, Orion B, and NGC~7538.  Much of this discrepancy can be attributed to
differences in analytical technique.  If we fit the \eight clumps in
NGC~7538, with a single power law, we find $\alpha = 0.4 \pm 0.2$, which
is consistent with the results obtained in the other studies which used a
single power law fit.  However, as can be seen in Figures \ref{fig:cumumf}
and \ref{fig:diffmf}, none of our mass functions would be fit well by a
single power law; there is a definite break point in the mass function,
whether physical or an artifact of incompleteness.  The mass functions of
\citet{tot}, \citet{kerton}, and \citet{mookerjea} all show evidence of a
similar break in the power law.  If the mass functions for each of these
regions were fit with a double power law, the derived exponent at the high
mass end should increase and might then agree with the other studies given
in Table \ref{tab:powerlaws}.  Especially because the origin of the break
is not well understood, we believe that a double power law whose break
point is a fitted parameter is the most appropriate way to fit these mass
functions.  This issue will be examined further in Paper~II.

	Another common technique for measuring the clump mass function in
star-forming regions is to extract the clumps from a CO spectral cube.  
Such studies typically find shallower mass functions than those which deal
solely with the mm/submm clumps.  \citet{kramer98} fitted single power law
mass functions to the clumps extracted from $^{13}$CO~(1$\rightarrow$0)
and C$^{18}$O (1$\rightarrow$0) maps of NGC~7538.  In the two maps, they
found $\gamma$ = 1.65 $\pm$ 0.05 and 1.79 $\pm$ 0.12, respectively, which
agree within the uncertainties with our result of 2.0~$\pm$~0.3 for the
high-mass end of the \eight clump mass function.  However, there appears
to be statistically significant disagreement between the mass functions
found in CO spectral line clump studies and mm/submm dust continuum
studies generally.  The mean slope of the high-mass end of the mm/submm
mass functions fitted with two power laws (Table \ref{tab:powerlaws}) is
$2.3 \pm 0.1$, whereas the mean slope of the CO mass functions for the
seven star-forming regions included in the study of \citet{kramer98} is
$1.69 \pm 0.02$.  The CO and dust continuum studies span roughly the same
range of clump masses: $\sim$10$^{-4}$--$10^{4}$ M$_{\odot}$ for the CO
clumps and $\sim$10$^{-2}$--$10^{3}$ M$_{\odot}$ for the dust continuum
studies.  Thus, the discrepancy between the average slopes of the high
mass ends of the CO and dust continuum mass functions appears to be real.  
An alternate interpretation is that the CO mass functions agree with the
dust continuum mass functions of the Lagoon Nebula, KR 140, and RCW 106,
which have a mean exponent of 1.6 $\pm$ 0.1.  However, we note that,
whereas the CO mass functions appear to be well fit by a single power law,
the dust continuum mass functions are clearly not.  If the high- and
low-mass portions of the dust continuum mass functions in these regions
were fitted separately, we believe the exponents of their high-mass power
laws would be found to agree with those quoted in Table
\ref{tab:powerlaws} for $\rho$ Oph, Orion B, and NGC~7538, and to disagree
with the exponents of the CO mass functions.

	A possible explanation for the apparently real discrepancy between
the CO spectral line and dust continuum mass functions is that the dust
maps trace denser, more likely gravitationally bound clumps than the CO
line maps, which may trace less dense, possibly transient structures
(\citealt{m01}, \citealt{tot}).  This hypothesis could be tested by
comparing CO and dust continuum maps of individual regions taken at
similar resolutions and with sensitivity to structure on similar scales.  
Repeating this analysis at multiple resolutions and for a variety of
regions would be very helpful.  Freeze-out of CO in cooler and/or less
massive clumps may also account for some of the discrepancy
\citep{mitchell01}.  Finally, some of the discrepancy may be attributable
to the recovery of different types of structures from spectral cubes
analyzed using algorithms like GAUSSCLUMPS \citep{sg90} or the 3D version
of clumpfind, compared to those recovered from integrated intensity maps
using algorithms like \clfind, as here.  To \clfind, every integrated
intensity peak represents a single clump, but the extra velocity
information available in a spectral cube can allow GAUSSCLUMPS to
decompose each peak into more than one clump along the line of sight,
potentially producing systematically different mass functions.

	What can we now conclude about the origin of the clump mass
function and what that might imply for the origin of the stellar IMF?  
The five studies summarized in Table~\ref{tab:powerlaws} cover three
different regions of star formation whose clump masses span five orders of
magnitude, from $\sim$0.03~M$_{\odot}$ clumps in \citet{dj2000b} to
$\sim$3000~M$_{\odot}$ clumps in this work.  In all cases, the high-mass
end of the mass function, which is not strongly affected by
incompleteness, is well-fit by a Salpeter-like power law with index
$\sim$2.0--2.5.  Similarly, the low-mass end of the mass function has a
power law index of $\sim$0.9--1.5 in all cases.  This agreement suggests
that the processes which shape the clump mass function are self-similar,
at least across five orders of magnitude in clump mass.  In other words,
the overall shape of the mass function is independent of the total mass of
the molecular cloud, over the range of cloud masses studied.  If there is
an intrinsic scale in a cloud's mass function, it appears to manifest as
the position of the break point between the low and high mass ends.  The
results from the low-mass star-forming regions summarized in Table
\ref{tab:powerlaws} suggest that the break point may occur at
$\sim$0.5--1.0 M$_{\odot}$, which roughly matches the break point in the
stellar IMF \citep{kroupa}.  However, the appearance of a break point in
the mass function of NGC~7538 suggests that this intrinsic scale of the
clump mass function may scale with some global property of the parent
cloud, such as its total mass or the number of Jeans masses present.  If
this is the case, then the mass function of clumps seen in high-mass
regions may not simply be an extension of the high-mass end of the mass
function seen in lower-mass regions.  Instead, it may be a scaled-up
version of the whole mass function seen in low-mass regions, including the
break point.  Indications of a similar high-mass break point can also be
seen in the dust continuum clump mass functions of the Lagoon nebula
\citep{tot} and the molecular cloud complex RCW 106 \citep{mookerjea}.  
However, in all of the dust continuum studies, the existence and/or the
position of a break point could be a function of incompleteness.  
Continuum surveys at higher resolutions and sensitivities will be required
to confirm the existence and position of the break point in the clump mass
function.

	Our observations of NGC~7538 are also consistent with theoretical
and numerical studies of molecular clouds in which turbulent fragmentation
sets the clump mass function, at least on large scales.  In their
simulations of several Jeans masses of self-gravitating, turbulent gas,
\citet{tilley} found that turbulent fragmentation alone is sufficient to
give rise to a Salpeter-like mass function with $\gamma = 2.29$.  A
similar result was obtained by \citet{gammie} for a magnetized, turbulent
gas, who found $\gamma \simeq 2.3$ for both gravitationally bound and
unbound clumps (and the ensemble of both).  \citet{pn02} used an
analytical approach to turbulent fragmentation to derive the same result
($\gamma = 2.33$).

	Our inability to estimate accurately the elapsed time since the
onset of collapse in any given molecular cloud complicates this line of
reasoning.  For example, we cannot be entirely confident that the
population of clumps in NGC~7538 is mostly unevolved; some coalescence of
clumps may already have taken place, and some fraction of each clump's
mass may already have accreted onto one or more embedded protostars.  The
presence of a few very young, massive outflows in only the largest clumps
and the absence of widespread maser activity suggests that star formation
in NGC~7538 is in its very early stages, but only deep infrared surveys
can address this issue adequately.  

	It is not yet clear what fraction of the clumps in NGC~7538 may
form more than one star, and with what distribution of masses.  This
complicates any attempt to extrapolate a connection between stellar-mass
clumps on linear scales of $\sim$ 0.02 pc and the clumps in our map, on
scales of $\sim$ 0.2~pc.  Nonetheless, we believe it is reasonable to
interpret the early emergence of a Salpeter-like clump mass spectrum, on
both large and small scales, as evidence that self-similar turbulent
fragmentation may be sufficient to determine the final distribution of
stellar masses, reducing the need to invoke other processes.

\section{Summary}
\label{sec:sum}

	We have mapped the NGC~7538 massive star-forming region with high
sensitivity at \eight and \four.  After processing the images to remove
spurious large-scale structure, we used \clfind to locate 77 clumps in the
\eight image and 67 in the \four image.  Our calibrations take into
account contamination from CO line emission, radio continuum emission, and
the error beam at both wavelengths.  Assuming a uniform dust temperature
of 35~K, we find that the masses of the \eight and \four clumps range from
1.2 to 2700 M$_{\odot}$ and 4 to 3100 M$_{\odot}$, respectively.  We find
a positive correlation between spectral index, $\alpha$, and submillimeter
emission.  This is the opposite result to that obtained by \citet{fms01},
but the difference is entirely accounted for by differences in the
techniques used to make the two maps.  The positive correlation between
high submm flux and high spectral index is likely due to the higher
temperatures at the centers of massive cores, particularly those
containing embedded protostars, but may also reflect spatial variations in
the dust emissivity index, $\beta$.  Using the spectral index map and the
assumption of a constant dust emissivity index, $\beta$, we derive
temperatures for 42 of the 77 clumps seen at \eight and use these in an
alternative calculation of the clump masses.

	Our most reliable analysis of the clump mass-radius relationship
gives ${\rm M}\propto {\rm R}^{x}$ with $x\sim 1.5-2.1$.  This result
suggests that the clumps are characterized by a nearly constant column
density, and that they probably have significant turbulent support.  The
similarity in the structures seen at \four and \eight, which differ in
resolution by a factor of two, suggests that the clumps we are seeing are 
coherent physical structures, and not superpositions of unrelated smaller 
objects.

	We have derived the first set of mass functions for a large number
of submm clumps in a single high-mass star-forming region.  At the
high-mass end where the data are not strongly affected by incompleteness,
the mass function is approximately Salpeter.  Our most reliable analysis
gives an exponent of 2.0 $\pm$ 0.3 (compared to 2.35 for the Salpeter IMF)
for the power law exponent of the high-mass end of the clump mass
function, covering the range $\sim100-3000$~M$_{\odot}$.  The submm
clump mass function of NGC~7538 agrees well with those found in $\rho$ Oph
and Orion B, using similar analytical techniques.  The range of clump
masses covered by these studies plus our own spans five orders of
magnitude.  We find that the difference in the power law exponents of the
CO and dust continuum clump mass functions is statistically significant.  
The dust continuum mass functions are typically steeper at the high-mass
end than their CO counterparts, though a definitive explanation for this
discrepancy is lacking.  We interpret the similar shapes of the clump mass
functions in both high-mass and low-mass star-forming regions as evidence
for the self-similarity of the physical processes which set the clump mass
function.  We suggest that turbulent fragmentation may be the dominant
process shaping the clump mass function in young star-forming regions.  
Finally, we interpret the early emergence of a Salpeter-like mass function
in several young star-forming regions as evidence that late-stage
processes such as competitive accretion and coalescence may not need to
play a significant role in the generation of the stellar IMF.

\begin{acknowledgements}

M.~A.~R. has been supported by an Ontario Graduate Scholarship in Science
and Technology.  Both M.~A.~R. and C.~D.~W. are supported by the Natural
Sciences and Engineering Research Council of Canada (NSERC).  We would
like to thank the referee for numerous useful suggestions.  M.~A.~R.  
would like to acknowledge several helpful conversations with Doug
Johnstone, David Tilley, Doug Welch, and Gerald Moriarty-Schieven.  The
James Clerk Maxwell Telescope is operated by The Joint Astronomy Centre on
behalf of the Particle Physics and Astronomy Research Council of the
United Kingdom, the Netherlands Organisation for Scientific Research, and
the National Research Council of Canada.  This publication makes use of
data products from the Two Micron All Sky Survey, which is a joint project
of the University of Massachusetts and the Infrared Processing and
Analysis Center/California Institute of Technology, funded by the National
Aeronautics and Space Administration and the National Science Foundation.

\end{acknowledgements}


\clearpage

\clearpage

\clearpage

\clearpage

\clearpage

\clearpage

\clearpage

\clearpage

\begin{deluxetable}{lccccc}
\tabletypesize{\footnotesize}
\tablecaption{Two-Gaussian Beam Fits\label{tab:beams}}
\tablewidth{0pt}
\tablehead{
 &  & \multicolumn{2}{c}{Primary Beam} & \multicolumn{2}{c}{First Error Beam} \\
 & Calibrator & \colhead{Relative Amplitude} & \colhead{FWHM (arcsec)} & \colhead{Relative Amplitude} & \colhead{FWHM (arcsec)}}
\startdata
\multirow{2}*{\eight}
 & Uranus & 0.95$\pm$0.08 & 14.2$\pm$0.5 & 0.05$\pm$0.03 & 30$\pm$5 \\
 & CRL 2688 & 0.9$\pm$0.1 & 15.1$\pm$0.9 & 0.04$\pm$0.04 & 40$\pm$10 \\ \tableline
\multirow{2}*{\four} 
 & Uranus & 0.9$\pm$0.1 & 8.5$\pm$0.6 & 0.06$\pm$0.03 & 27$\pm$4 \\
 & CRL 2688 & 0.9$\pm$0.2 & 9.2$\pm$0.7 & 0.08$\pm$0.04 & 30$\pm$5 \\
\enddata
\end{deluxetable}

\clearpage

\begin{deluxetable}{ccccccccccc}
\tabletypesize{\tiny}
\tablecaption{Properties of the \eight Clumps\label{tab:850clumps}}
\tablewidth{0pt}
\tablehead{
\colhead{Name} & \colhead{R.A.}  & \colhead{Dec.} & \colhead{R$_{\rm eff}$} & 
\colhead{$S_{\rm peak}$\tablenotemark{a}} & \colhead{$S^{\rm int}_{850}$\tablenotemark{a}} & 
\colhead{$\langle{\alpha}\rangle$\tablenotemark{b}} & 
\colhead{$\langle {\rm T}_{d}\rangle$\tablenotemark{c}} & 
\colhead{M$_{{\rm T}_{d}=35K}$\tablenotemark{a}} & 
\colhead{M$_{{\rm T}_{d}}$\tablenotemark{a}} & \colhead{n$_{\rm psc}$\tablenotemark{d}} \\
\colhead{(NGC7538-)} & \colhead{(J2000)} & \colhead{(J2000)} & \colhead{(pc)} & \colhead{(Jy beam$^{-1}$)} & 
\colhead{(Jy)} & \colhead{} &  
\colhead{(K)} & \colhead{(M$_{\odot}$)} & \colhead{(M$_{\odot}$)} & \colhead{}}
\startdata
                     SMM1 & 23 12 49.8 & +61 27 28 &  0.17 &   0.22$\pm$0.01 &   0.38$\pm$0.02 &            2.55 &              16 &    15.8$\pm$0.7 &        50$\pm$2 &   1\\
                     SMM2 & 23 12 52.1 & +61 25 56 &  0.03 &   0.10$\pm$0.01 &   0.06$\pm$0.01 &            3.99 &         \nodata &     2.4$\pm$0.5 &         \nodata &   0\\
                     SMM3 & 23 12 53.2 & +61 27 42 &  0.16 &   0.37$\pm$0.01 &   0.58$\pm$0.08 &            3.01 &              28 &        24$\pm$3 &        32$\pm$4 &   0\\
                     SMM4 & 23 12 54.6 & +61 27 00 &  0.28 &   0.78$\pm$0.03 &     2.1$\pm$0.3 &            3.33 &              72 &       90$\pm$10 &        37$\pm$5 &   3\\
                     SMM5 & 23 12 55.1 & +61 25 48 &  0.05 &   0.16$\pm$0.01 &   0.13$\pm$0.01 &            3.66 &         \nodata &     5.2$\pm$0.5 &         \nodata &   1\\
                     SMM6 & 23 12 55.4 & +61 27 32 &  0.08 &   0.42$\pm$0.02 &   0.68$\pm$0.06 &            3.18 &              40 &        28$\pm$2 &        24$\pm$2 &   1\\
                     SMM7 & 23 12 56.2 & +61 26 20 &  0.15 &   0.18$\pm$0.01 &   0.27$\pm$0.01 &            3.59 &         \nodata &    11.2$\pm$0.5 &         \nodata &   1\\
                     SMM8 & 23 12 56.8 & +61 25 54 &  0.08 &   0.19$\pm$0.01 &   0.20$\pm$0.01 &            3.67 &         \nodata &     8.4$\pm$0.3 &         \nodata &   2\\
                     SMM9 & 23 12 58.7 & +61 27 20 &  0.06 &   0.54$\pm$0.02 &   0.83$\pm$0.07 &            3.44 &             204 &        35$\pm$3 &     4.8$\pm$0.4 &   3\\
                    SMM10 & 23 12 59.0 & +61 27 52 &  0.20 &   0.63$\pm$0.03 &     1.2$\pm$0.1 &            3.56 &         \nodata &        51$\pm$6 &         \nodata &   2\\
                    SMM11 & 23 12 59.3 & +61 27 30 &  0.10 &   0.56$\pm$0.02 &   0.67$\pm$0.05 &            3.54 &         \nodata &        28$\pm$2 &         \nodata &   1\\
                    SMM12 & 23 13 00.4 & +61 24 56 &  0.04 &   0.10$\pm$0.01 &   0.06$\pm$0.01 &         \nodata &         \nodata &     2.7$\pm$0.5 &         \nodata &   0\\
                    SMM13 & 23 13 03.4 & +61 27 52 &  0.12 &   0.17$\pm$0.01 &   0.20$\pm$0.04 &            3.50 &         \nodata &         8$\pm$2 &         \nodata &   0\\
                    SMM14 & 23 13 03.7 & +61 24 28 &  0.12 &   0.16$\pm$0.01 &   0.19$\pm$0.04 &         \nodata &         \nodata &         8$\pm$2 &         \nodata &   1\\
                    SMM15 & 23 13 05.4 & +61 24 50 &  0.06 &   0.11$\pm$0.01 &   0.04$\pm$0.03 &         \nodata &         \nodata &         2$\pm$1 &         \nodata &   0\\
                    SMM16 & 23 13 05.7 & +61 28 04 &  0.14 &   0.25$\pm$0.01 &   0.39$\pm$0.01 &            3.45 &             219 &    16.2$\pm$0.6 &   2.09$\pm$0.07 &   1\\
                    SMM17 & 23 13 09.0 & +61 23 56 &  0.13 &   0.17$\pm$0.01 &   0.23$\pm$0.01 &            3.78 &         \nodata &     9.5$\pm$0.4 &         \nodata &   0\\
                    SMM18 & 23 13 10.1 & +61 28 18 &  0.34 &   1.09$\pm$0.04 &     3.3$\pm$0.5 &            3.49 &         \nodata &      140$\pm$20 &         \nodata &   3\\
                    SMM19 & 23 13 15.7 & +61 29 10 &  0.26 &   0.85$\pm$0.03 &     2.1$\pm$0.3 &            3.13 &              35 &       90$\pm$10 &       90$\pm$10 &   7\\
                    SMM20 & 23 13 18.4 & +61 29 28 &  0.16 &   0.39$\pm$0.02 &   0.73$\pm$0.03 &            3.06 &              30 &        31$\pm$1 &        37$\pm$1 &   1\\
                    SMM21 & 23 13 18.4 & +61 28 50 &  0.25 &   0.25$\pm$0.01 &   0.70$\pm$0.03 &            2.64 &              17 &        29$\pm$1 &        81$\pm$4 &   2\\
                    SMM22 & 23 13 19.8 & +61 25 20 &  0.27 &   1.05$\pm$0.04 &     3.0$\pm$0.4 &            3.13 &              35 &      120$\pm$20 &      120$\pm$20 &   6\\
                    SMM23 & 23 13 22.0 & +61 24 54 &  0.28 &   0.84$\pm$0.03 &     2.6$\pm$0.3 &            3.53 &         \nodata &      110$\pm$10 &         \nodata &   0\\
                    SMM24 & 23 13 22.3 & +61 29 12 &  0.22 &   0.32$\pm$0.01 &   0.91$\pm$0.04 &            2.57 &              16 &        38$\pm$2 &       116$\pm$6 &   3\\
                    SMM25 & 23 13 23.4 & +61 31 06 &  0.06 &   0.12$\pm$0.01 &   0.09$\pm$0.01 &         \nodata &         \nodata &     3.7$\pm$0.3 &         \nodata &   2\\
                    SMM26 & 23 13 24.5 & +61 25 28 &  0.21 &   0.54$\pm$0.02 &     1.6$\pm$0.2 &            3.20 &              42 &        69$\pm$8 &        54$\pm$6 &   1\\
                    SMM27 & 23 13 25.1 & +61 25 00 &  0.17 &   0.76$\pm$0.03 &     1.9$\pm$0.2 &            3.59 &         \nodata &        81$\pm$8 &         \nodata &   1\\
   SMM28\tablenotemark{e} & 23 13 25.6 & +61 29 52 &  0.14 &   0.56$\pm$0.02 &     1.1$\pm$0.1 &            2.57 &              17 &        43$\pm$4 &      120$\pm$10 &   3\\
                    SMM29 & 23 13 25.6 & +61 24 18 &  0.23 &   0.14$\pm$0.01 &   0.35$\pm$0.04 &            3.68 &         \nodata &        15$\pm$1 &         \nodata &   1\\
   SMM30\tablenotemark{e} & 23 13 26.8 & +61 29 26 &  0.25 &   1.05$\pm$0.04 &     3.5$\pm$0.4 &            3.00 &              29 &      140$\pm$20 &      180$\pm$20 &   8\\
   SMM31\tablenotemark{e} & 23 13 27.0 & +61 30 12 &  0.19 &   0.44$\pm$0.02 &   0.85$\pm$0.09 &            2.46 &              15 &        34$\pm$4 &      110$\pm$10 &   8\\
                    SMM32 & 23 13 28.4 & +61 25 58 &  0.16 &   0.18$\pm$0.01 &   0.30$\pm$0.02 &            2.95 &              25 &    12.4$\pm$0.9 &        19$\pm$1 &   0\\
                    SMM33 & 23 13 29.0 & +61 25 28 &  0.17 &   0.72$\pm$0.03 &     1.9$\pm$0.2 &            3.41 &             128 &        78$\pm$8 &        18$\pm$2 &   0\\
   SMM34\tablenotemark{e} & 23 13 29.5 & +61 30 04 &  0.23 &   0.77$\pm$0.02 &     1.7$\pm$0.2 &            2.32 &              17 &        57$\pm$8 &      150$\pm$20 &  10\\
                    SMM35 & 23 13 29.8 & +61 25 04 &  0.20 &   0.53$\pm$0.02 &     1.3$\pm$0.1 &            3.62 &         \nodata &        53$\pm$6 &         \nodata &   2\\
   SMM36\tablenotemark{e} & 23 13 30.1 & +61 30 26 &  0.25 &   0.82$\pm$0.03 &     1.8$\pm$0.2 &            2.63 &              19 &        68$\pm$9 &      160$\pm$20 &   6\\
   SMM37\tablenotemark{e} & 23 13 30.6 & +61 29 02 &  0.31 &   1.99$\pm$0.08 &     6.2$\pm$0.8 &            3.30 &              68 &      250$\pm$30 &      120$\pm$20 &  13\\
   SMM38\tablenotemark{e} & 23 13 31.5 & +61 31 18 &  0.17 &   0.12$\pm$0.01 &   0.20$\pm$0.01 &         \nodata &         \nodata &     6.6$\pm$0.3 &         \nodata &   1\\
                    SMM39 & 23 13 32.0 & +61 26 56 &  0.16 &   0.15$\pm$0.01 &   0.24$\pm$0.01 &         \nodata &         \nodata &     9.9$\pm$0.2 &         \nodata &   0\\
                    SMM40 & 23 13 33.7 & +61 25 46 &  0.40 &   1.20$\pm$0.05 &     5.2$\pm$0.7 &            3.43 &             172 &      220$\pm$30 &        36$\pm$5 &   6\\
                    SMM41 & 23 13 33.7 & +61 31 52 &  0.19 &   0.19$\pm$0.01 &   0.36$\pm$0.02 &            2.17 &              12 &    15.0$\pm$0.7 &        77$\pm$3 &   0\\
                    SMM42 & 23 13 34.2 & +61 28 32 &  0.52 &   1.06$\pm$0.04 &     6.1$\pm$0.9 &            3.06 &              30 &      250$\pm$40 &      300$\pm$40 &  16\\
                    SMM43 & 23 13 35.6 & +61 29 28 &  0.09 &   0.09$\pm$0.01 &   0.08$\pm$0.01 &         \nodata &         \nodata &     3.3$\pm$0.2 &         \nodata &   3\\
                    SMM44 & 23 13 42.3 & +61 31 36 &  0.21 &   0.54$\pm$0.02 &     1.0$\pm$0.1 &            2.97 &              26 &        43$\pm$5 &        63$\pm$7 &   0\\
                    SMM45 & 23 13 43.7 & +61 31 34 &  0.23 &   0.54$\pm$0.02 &     0.9$\pm$0.1 &            3.11 &              34 &        38$\pm$5 &        39$\pm$5 &   0\\
                    SMM46 & 23 13 44.5 & +61 26 48 &  0.67 &    14.1$\pm$0.6 &        53$\pm$8 &            3.50 &         \nodata &    2200$\pm$300 &         \nodata &   4\\
   SMM47\tablenotemark{e} & 23 13 44.8 & +61 29 12 &  0.18 &   0.20$\pm$0.01 &   0.29$\pm$0.03 &            3.28 &             329 &        11$\pm$1 &     0.9$\pm$0.1 &   0\\
   SMM48\tablenotemark{e} & 23 13 45.3 & +61 28 10 &  0.68 &    18.1$\pm$0.7 &       70$\pm$10 &            3.44 &         \nodata &    2700$\pm$400 &         \nodata &   8\\
                    SMM49 & 23 13 45.3 & +61 25 50 &  0.14 &   0.21$\pm$0.01 &   0.33$\pm$0.01 &            1.48 &               8 &    13.6$\pm$0.4 &       143$\pm$4 &   1\\
                    SMM50 & 23 13 48.4 & +61 25 50 &  0.23 &   0.20$\pm$0.01 &   0.54$\pm$0.01 &            2.16 &              12 &    22.5$\pm$0.5 &       118$\pm$2 &   3\\
                    SMM51 & 23 13 49.2 & +61 29 10 &  0.18 &   0.16$\pm$0.01 &   0.31$\pm$0.08 &            2.66 &              17 &        13$\pm$4 &        35$\pm$9 &   4\\
                    SMM52 & 23 13 53.4 & +61 27 00 &  0.37 &   0.96$\pm$0.04 &     4.2$\pm$0.6 &            3.31 &              66 &      180$\pm$20 &       80$\pm$10 &   9\\
                    SMM53 & 23 13 53.9 & +61 28 52 &  0.06 &   0.10$\pm$0.01 &   0.03$\pm$0.01 &         \nodata &         \nodata &     1.4$\pm$0.3 &         \nodata &   0\\
                    SMM54 & 23 13 56.4 & +61 27 08 &  0.30 &   0.81$\pm$0.03 &     2.6$\pm$0.3 &            3.37 &              93 &      110$\pm$10 &        35$\pm$5 &  15\\
                    SMM55 & 23 13 56.7 & +61 27 50 &  0.17 &   0.15$\pm$0.01 &   0.24$\pm$0.01 &         \nodata &         \nodata &    10.0$\pm$0.2 &         \nodata &   0\\
                    SMM56 & 23 13 58.9 & +61 31 28 &  0.37 &   0.68$\pm$0.03 &     2.6$\pm$0.4 &            3.08 &              32 &      110$\pm$10 &      120$\pm$20 &   5\\
                    SMM57 & 23 13 59.5 & +61 30 54 &  0.18 &   1.25$\pm$0.05 &     2.6$\pm$0.3 &            3.29 &              59 &      110$\pm$10 &        59$\pm$7 &   3\\
                    SMM58 & 23 14 00.6 & +61 28 02 &  0.06 &   0.11$\pm$0.01 &   0.07$\pm$0.01 &         \nodata &         \nodata &     2.8$\pm$0.3 &         \nodata &   1\\
                    SMM59 & 23 14 00.9 & +61 29 50 &  0.04 &   0.09$\pm$0.01 &   0.05$\pm$0.01 &         \nodata &         \nodata &     2.1$\pm$0.2 &         \nodata &   0\\
                    SMM60 & 23 14 02.0 & +61 27 20 &  0.59 &     3.9$\pm$0.2 &        14$\pm$2 &            3.45 &             225 &      600$\pm$90 &       80$\pm$10 &   4\\
                    SMM61 & 23 14 02.0 & +61 30 16 &  0.14 &   0.81$\pm$0.03 &     1.0$\pm$0.1 &            3.24 &              48 &        41$\pm$4 &        27$\pm$3 &   4\\
                    SMM62 & 23 14 02.3 & +61 29 34 &  0.06 &   0.12$\pm$0.01 &   0.10$\pm$0.01 &         \nodata &         \nodata &     4.0$\pm$0.2 &         \nodata &   0\\
                    SMM63 & 23 14 05.9 & +61 31 00 &  0.26 &   1.45$\pm$0.06 &     3.5$\pm$0.5 &            3.32 &              69 &      140$\pm$20 &        64$\pm$8 &   1\\
                    SMM64 & 23 14 05.9 & +61 30 10 &  0.16 &   0.34$\pm$0.01 &   0.69$\pm$0.03 &            3.18 &              40 &        29$\pm$1 &    24.4$\pm$0.9 &   1\\
                    SMM65 & 23 14 06.2 & +61 31 30 &  0.27 &   0.75$\pm$0.03 &     2.0$\pm$0.2 &            3.24 &              49 &       80$\pm$10 &        54$\pm$7 &   3\\
                    SMM66 & 23 14 08.4 & +61 29 44 &  0.20 &   0.33$\pm$0.01 &   0.71$\pm$0.05 &            3.15 &              37 &        29$\pm$2 &        28$\pm$2 &   1\\
                    SMM67 & 23 14 08.4 & +61 30 00 &  0.10 &   0.29$\pm$0.01 &   0.36$\pm$0.07 &            3.10 &              33 &        15$\pm$3 &        16$\pm$3 &   1\\
                    SMM68 & 23 14 11.7 & +61 31 46 &  0.15 &   0.12$\pm$0.01 &   0.20$\pm$0.01 &         \nodata &         \nodata &   8.41$\pm$0.08 &         \nodata &   2\\
                    SMM69 & 23 14 12.3 & +61 26 14 &  0.06 &   0.09$\pm$0.01 &   0.03$\pm$0.01 &         \nodata &         \nodata &     1.4$\pm$0.3 &         \nodata &   0\\
                    SMM70 & 23 14 15.0 & +61 25 54 &  0.06 &   0.13$\pm$0.01 &   0.10$\pm$0.01 &            3.20 &              43 &     4.1$\pm$0.3 &     3.1$\pm$0.2 &   1\\
                    SMM71 & 23 14 18.1 & +61 26 00 &  0.08 &   0.13$\pm$0.01 &   0.12$\pm$0.01 &         \nodata &         \nodata &     4.9$\pm$0.3 &         \nodata &   1\\
                    SMM72 & 23 14 22.8 & +61 29 50 &  0.06 &   0.11$\pm$0.01 &   0.03$\pm$0.01 &         \nodata &         \nodata &     1.2$\pm$0.2 &         \nodata &   0\\
                    SMM73 & 23 14 26.7 & +61 31 34 &  0.06 &   0.09$\pm$0.01 &   0.03$\pm$0.01 &         \nodata &         \nodata &     1.3$\pm$0.3 &         \nodata &   0\\
                    SMM74 & 23 14 29.2 & +61 31 30 &  0.06 &   0.11$\pm$0.01 &   0.08$\pm$0.01 &         \nodata &         \nodata &   3.13$\pm$0.07 &         \nodata &   0\\
                    SMM75 & 23 14 29.7 & +61 28 52 &  0.09 &   0.15$\pm$0.01 &   0.14$\pm$0.01 &            4.12 &         \nodata &     5.9$\pm$0.1 &         \nodata &   0\\
                    SMM76 & 23 14 32.5 & +61 31 14 &  0.06 &   0.10$\pm$0.01 &   0.06$\pm$0.01 &         \nodata &         \nodata &     2.3$\pm$0.6 &         \nodata &   2\\
                    SMM77 & 23 14 36.7 & +61 27 46 &  0.07 &   0.13$\pm$0.01 &   0.11$\pm$0.01 &            3.24 &              49 &     4.5$\pm$0.2 &     3.0$\pm$0.1 &   0\\
\enddata
\tablenotetext{a}{The uncertainties stated in this table are the systematic ones, composed
of the uncertainties in the gain calibration, the sky opacities, and the corrections due to
the error beam.  The systematic uncertainties are typically 
significantly larger than the random errors.  The exception is the peak flux, where the random
error of 
$\sigma = 0.021$ Jy beam$^{-1}$ dominates the systematic error for the lower peak fluxes.}
\tablenotetext{b}{The systematic uncertainty in the spectral index, 
$\alpha$, is 13\%.}
\tablenotetext{c}{See Section \ref{sec:tvar} for a discussion of the uncertainties in the
temperatures.  Temperatures are omitted where high spectral index makes them incalculable, or where 
no reliable spectral index can be calculated (see text).}
\tablenotetext{d}{Number of 2MASS point sources contained within the clump's 0.5$S_{\rm peak}$ contour.}
\tablenotetext{e}{Denotes a clump to which corrections for free-free emission have 
been applied in the calculation of the spectral index, dust temperature, and 
masses.  The free-free correction has \emph{not} been applied to 
the peak and integrated fluxes listed here.}
\end{deluxetable}

\clearpage

\begin{deluxetable}{cccccccc}
\tabletypesize{\scriptsize}
\tablecaption{Properties of the \four Clumps\label{tab:450clumps}}
\tablewidth{0pt}
\tablehead{
\colhead{Name\tablenotemark{a}} & \colhead{R.A.}  & \colhead{Dec.} & 
\colhead{R$_{\rm eff}$} & 
\colhead{$S_{\rm peak}$\tablenotemark{b}} & \colhead{$S^{\rm 
int}_{450}$\tablenotemark{b}} & 
\colhead{M$_{{\rm T}_{d}=35K}$\tablenotemark{b}} & \colhead{n$_{\rm psc}$\tablenotemark{c}} \\
\colhead{(NGC7538-)} & \colhead{(J2000)} & \colhead{(J2000)} & 
\colhead{(pc)} & \colhead{Jy beam$^{-1}$} & \colhead{(Jy)} & \colhead{(M$_{\odot}$)} & \colhead{} }
\startdata
                    SMM4A & 23 12 52.6 & +61 26 36 &  0.13 &     1.7$\pm$0.2 &     3.7$\pm$0.3 &        15$\pm$1 &   0 \\
                    SMM3A & 23 12 53.7 & +61 27 44 &  0.10 &     1.5$\pm$0.2 &     2.6$\pm$0.3 &        11$\pm$1 &   0 \\
                    SMM4B & 23 12 55.1 & +61 26 58 &  0.24 &     3.6$\pm$0.4 &        22$\pm$2 &        93$\pm$7 &   2 \\
                    SMM6A & 23 12 55.1 & +61 27 34 &  0.14 &     2.0$\pm$0.2 &         5$\pm$1 &        22$\pm$6 &   1 \\
                    SMM8A & 23 12 56.2 & +61 25 52 &  0.12 &     1.2$\pm$0.1 &   2.28$\pm$0.01 &   9.49$\pm$0.05 &   1 \\
                    SMM9A & 23 12 57.1 & +61 27 14 &  0.17 &     2.3$\pm$0.3 &     8.8$\pm$0.4 &        37$\pm$2 &   3 \\
                   SMM10A & 23 12 59.3 & +61 27 50 &  0.27 &     3.2$\pm$0.4 &        19$\pm$3 &       80$\pm$10 &   1 \\
                    SMM9B & 23 12 59.3 & +61 27 16 &  0.16 &     2.5$\pm$0.3 &         8$\pm$1 &        33$\pm$5 &   2 \\
                   SMM16A & 23 13 05.7 & +61 27 54 &  0.18 &     1.2$\pm$0.1 &     4.7$\pm$0.4 &        19$\pm$2 &   1 \\
                   SMM18A & 23 13 10.1 & +61 28 22 &  0.34 &     5.7$\pm$0.7 &        35$\pm$6 &      150$\pm$30 &   2 \\
                   SMM19A & 23 13 15.9 & +61 29 10 &  0.26 &     4.0$\pm$0.5 &        18$\pm$3 &       70$\pm$10 &   4 \\
                   SMM20A & 23 13 18.7 & +61 29 28 &  0.20 &     1.8$\pm$0.2 &     7.7$\pm$0.7 &        32$\pm$3 &   1 \\
                   SMM22B & 23 13 20.4 & +61 25 24 &  0.23 &     4.5$\pm$0.5 &        18$\pm$3 &       80$\pm$10 &   5 \\
                   SMM21A & 23 13 20.4 & +61 28 44 &  0.13 &     1.1$\pm$0.1 &   2.46$\pm$0.01 &  10.24$\pm$0.04 &   0 \\
                   SMM22A & 23 13 21.2 & +61 25 12 &  0.19 &     3.7$\pm$0.4 &        14$\pm$2 &        59$\pm$8 &   1 \\
                   SMM23A & 23 13 22.6 & +61 24 56 &  0.19 &     3.7$\pm$0.4 &        17$\pm$2 &        69$\pm$9 &   0 \\
                   SMM26C & 23 13 23.7 & +61 25 58 &  0.07 &     1.0$\pm$0.1 &     1.0$\pm$0.4 &         4$\pm$2 &   1 \\
                   SMM26A & 23 13 24.3 & +61 25 20 &  0.15 &     2.5$\pm$0.3 &        10$\pm$2 &       40$\pm$10 &   0 \\
                   SMM26B & 23 13 24.8 & +61 25 38 &  0.21 &     2.2$\pm$0.3 &     9.8$\pm$0.4 &        41$\pm$2 &   1 \\
                   SMM27B & 23 13 25.4 & +61 25 02 &  0.19 &     3.4$\pm$0.4 &        18$\pm$2 &       80$\pm$10 &   1 \\
                   SMM30A & 23 13 25.6 & +61 29 14 &  0.18 &     2.5$\pm$0.3 &         8$\pm$1 &        34$\pm$4 &   0 \\
                   SMM28A & 23 13 26.2 & +61 29 50 &  0.17 &     1.5$\pm$0.2 &     5.7$\pm$0.8 &        24$\pm$3 &   1 \\
                   SMM28B & 23 13 26.5 & +61 30 06 &  0.17 &     1.6$\pm$0.2 &     5.0$\pm$0.3 &        21$\pm$1 &   6 \\
                   SMM30B & 23 13 27.3 & +61 29 26 &  0.18 &     4.4$\pm$0.5 &        16$\pm$2 &        68$\pm$9 &   5 \\
                   SMM37A & 23 13 27.6 & +61 29 00 &  0.22 &     5.0$\pm$0.6 &        24$\pm$4 &      100$\pm$20 &   2 \\
                   SMM27A & 23 13 27.6 & +61 24 40 &  0.14 &     1.5$\pm$0.2 &     4.0$\pm$0.6 &        17$\pm$2 &   0 \\
                   SMM33A & 23 13 29.5 & +61 25 26 &  0.26 &     3.4$\pm$0.4 &        23$\pm$4 &       90$\pm$20 &   0 \\
                   SMM34B & 23 13 29.5 & +61 30 04 &  0.15 &     2.3$\pm$0.3 &         5$\pm$2 &        23$\pm$7 &   4 \\
                   SMM34A & 23 13 29.5 & +61 29 40 &  0.09 &     0.9$\pm$0.1 &   1.29$\pm$0.03 &     5.4$\pm$0.1 &   1 \\
                   SMM34C & 23 13 30.1 & +61 30 14 &  0.13 &     1.9$\pm$0.2 &         4$\pm$1 &        17$\pm$5 &   4 \\
                   SMM36B & 23 13 30.1 & +61 30 46 &  0.08 &     0.9$\pm$0.1 &   1.14$\pm$0.02 &   4.73$\pm$0.07 &   2 \\
                   SMM36A & 23 13 30.4 & +61 30 28 &  0.15 &     3.3$\pm$0.4 &         8$\pm$1 &        34$\pm$4 &   3 \\
                   SMM37B & 23 13 30.9 & +61 29 02 &  0.25 &        10$\pm$1 &        47$\pm$8 &      200$\pm$30 &   5 \\
                   SMM42B & 23 13 31.5 & +61 28 18 &  0.16 &     1.8$\pm$0.2 &     5.9$\pm$0.5 &        24$\pm$2 &   1 \\
                   SMM35A & 23 13 32.0 & +61 25 00 &  0.19 &     2.1$\pm$0.3 &     9.2$\pm$0.7 &        38$\pm$3 &   3 \\
                   SMM40A & 23 13 33.7 & +61 25 42 &  0.42 &     5.9$\pm$0.7 &       60$\pm$10 &      230$\pm$40 &   4 \\
                   SMM42C & 23 13 34.0 & +61 28 30 &  0.40 &     4.5$\pm$0.5 &        44$\pm$8 &      180$\pm$30 &  13 \\
                   SMM42A & 23 13 34.0 & +61 27 50 &  0.13 &     0.9$\pm$0.1 &   2.16$\pm$0.03 &     9.0$\pm$0.1 &   0 \\
                   SMM44A & 23 13 39.8 & +61 31 22 &  0.11 &     0.8$\pm$0.1 &     1.8$\pm$0.2 &     7.3$\pm$0.8 &   0 \\
                   SMM45A & 23 13 43.7 & +61 31 38 &  0.23 &     2.4$\pm$0.3 &    10.9$\pm$0.6 &        45$\pm$2 &   0 \\
                   SMM46A & 23 13 44.8 & +61 26 48 &  0.63 &       80$\pm$10 &     500$\pm$100 &    2300$\pm$400 &   2 \\
  SMM48A\tablenotemark{d} & 23 13 45.3 & +61 28 10 &  0.64 &        73$\pm$8 &     800$\pm$100 &    3000$\pm$600 &   6 \\
                   SMM47A & 23 13 45.3 & +61 29 14 &  0.11 &     1.0$\pm$0.1 &   2.04$\pm$0.03 &     8.5$\pm$0.1 &   1 \\
                   SMM48C & 23 13 47.0 & +61 28 54 &  0.16 &     1.3$\pm$0.2 &     4.2$\pm$0.7 &        17$\pm$3 &   3 \\
                   SMM46B & 23 13 49.8 & +61 26 58 &  0.35 &        11$\pm$1 &       80$\pm$10 &      310$\pm$60 &   1 \\
                   SMM48B & 23 13 51.2 & +61 28 48 &  0.18 &     1.2$\pm$0.1 &   4.70$\pm$0.02 &  19.53$\pm$0.08 &   1 \\
                   SMM56A & 23 13 51.4 & +61 31 22 &  0.08 &     0.9$\pm$0.1 &     1.2$\pm$0.5 &         5$\pm$2 &   0 \\
                   SMM56C & 23 13 53.7 & +61 31 26 &  0.11 &     1.3$\pm$0.2 &   2.39$\pm$0.04 &     9.9$\pm$0.2 &   0 \\
                   SMM52B & 23 13 54.2 & +61 27 00 &  0.17 &     5.1$\pm$0.6 &        19$\pm$3 &       80$\pm$10 &   4 \\
                   SMM52A & 23 13 54.5 & +61 26 50 &  0.27 &     4.9$\pm$0.6 &        24$\pm$4 &      100$\pm$20 &   7 \\
                   SMM56D & 23 13 55.1 & +61 31 32 &  0.10 &     1.6$\pm$0.2 &   3.06$\pm$0.02 &  12.75$\pm$0.07 &   1 \\
                   SMM54A & 23 13 56.4 & +61 27 12 &  0.23 &     4.2$\pm$0.5 &        18$\pm$3 &       70$\pm$10 &  11 \\
                   SMM56E & 23 13 56.7 & +61 31 36 &  0.13 &     2.4$\pm$0.3 &         5$\pm$1 &        22$\pm$5 &   0 \\
                   SMM57A & 23 13 59.8 & +61 30 52 &  0.24 &     6.6$\pm$0.8 &        25$\pm$4 &      100$\pm$20 &   0 \\
                   SMM56B & 23 13 59.8 & +61 31 26 &  0.24 &     2.8$\pm$0.3 &        15$\pm$1 &        62$\pm$4 &   5 \\
                   SMM60B & 23 14 00.9 & +61 27 02 &  0.25 &         8$\pm$1 &        42$\pm$7 &      180$\pm$30 &  10 \\
                   SMM60D & 23 14 02.0 & +61 27 20 &  0.26 &        27$\pm$3 &       90$\pm$10 &      360$\pm$60 &   2 \\
                   SMM61A & 23 14 02.3 & +61 30 16 &  0.17 &     4.6$\pm$0.6 &         9$\pm$1 &        37$\pm$5 &   2 \\
                   SMM60A & 23 14 05.0 & +61 26 58 &  0.32 &     6.5$\pm$0.8 &        39$\pm$7 &      160$\pm$30 &   1 \\
                   SMM64A & 23 14 05.6 & +61 30 12 &  0.18 &     1.7$\pm$0.2 &     5.9$\pm$0.4 &        24$\pm$2 &   0 \\
                   SMM65A & 23 14 05.9 & +61 31 28 &  0.27 &     3.3$\pm$0.4 &        17$\pm$3 &       70$\pm$10 &   1 \\
                   SMM63A & 23 14 06.2 & +61 31 00 &  0.30 &     7.1$\pm$0.9 &        33$\pm$6 &      140$\pm$20 &   1 \\
                   SMM66A & 23 14 09.2 & +61 29 48 &  0.15 &     1.3$\pm$0.2 &   3.88$\pm$0.03 &    16.1$\pm$0.1 &   0 \\
                   SMM67A & 23 14 09.5 & +61 30 06 &  0.19 &     1.2$\pm$0.1 &   5.57$\pm$0.06 &    23.1$\pm$0.2 &   0 \\
                   SMM60C & 23 14 11.4 & +61 27 18 &  0.15 &     1.2$\pm$0.1 &     3.3$\pm$0.2 &    13.6$\pm$0.9 &   0 \\
                   SMM75A & 23 14 30.0 & +61 29 00 &  0.12 &     1.2$\pm$0.1 &     2.5$\pm$0.4 &        11$\pm$2 &   0 \\
                   SMM77A & 23 14 36.4 & +61 27 44 &  0.11 &     0.9$\pm$0.1 &     1.6$\pm$0.2 &     6.7$\pm$0.8 &   0 \\
\enddata
\tablenotetext{a}{The names of the \four clumps have been set to reflect the names of the \eight
clumps in which their peaks appear.  Thus, clumps SMM26A-C are the three \four clumps whose peaks
appear within the boundaries of \eight clump SMM26.}
\tablenotetext{b}{The uncertainties stated in this table are the systematic ones, composed of the uncertainties in the gain calibration,
the sky opacities, and the corrections due to the error beam.  The systematic uncertainties are typically significantly larger than the
random errors.  The exception is the peak flux, where the random error of $\sigma = 0.18$ Jy beam$^{-1}$ dominates the systematic error
for the lower peak fluxes.}
\tablenotetext{c}{Number of 2MASS point sources contained within the clump's 0.5$S_{\rm peak}$ contour.}
\tablenotetext{d}{Denotes a clump to which corrections for free-free emission have 
been applied in the calculation of the spectral index, dust temperature, and 
masses.  The free-free correction has \emph{not} been applied to 
the peak and integrated fluxes listed here.}
\end{deluxetable}

\clearpage

\begin{deluxetable}{cc}
\tablecaption{Submm Clumps Associated with IRS Sources\label{tab:assoc}}
\tablewidth{0pt}
\tablehead{
\colhead{IRS Source} & \colhead{Associated} \\
\colhead{Number} & \colhead{Submm Clump}
}
\startdata
1, 2, \& 3 & SMM48 \\
4 & SMM37 \\
5 & SMM34 \\
9 & SMM60 \\
11 & SMM46 \\
\enddata
\end{deluxetable}

\clearpage

\begin{deluxetable}{ccccccccccc}
\tabletypesize{\tiny}
\tablecaption{Power Law Fits to (Sub)Millimeter Clump Mass 
Functions\label{tab:powerlaws}}
\tablewidth{0pt}
\tablehead{
 &\colhead{$\lambda$} & \colhead{HPBW} &\colhead{Fitted Mass} & & \colhead{M$_{\rm break}$} & 
\multicolumn{2}{c}{CMF 
Exponents\tablenotemark{b}} & 
\multicolumn{2}{c}{DMF Exponents\tablenotemark{b}} & \\ \cline{7-8} \cline{9-10}
\colhead{Region} &\colhead{($\mu{\rm m}$)} & \colhead{(pc)} & \colhead{Range (M$_{\odot}$)} & 
\colhead{N$_{\rm 
fit}$\tablenotemark{a}} & 
\colhead{(M$_{\odot}$)} & 
\colhead{$\alpha_{\rm low}$} & \colhead{$\alpha_{\rm high}$} 
&\colhead{$\gamma_{\rm low}$} & \colhead{$\gamma_{\rm high}$} & \colhead{Ref.}
}
\startdata
\multicolumn{10}{c}{Double Power Law Fits} \\ \tableline
NGC~7538 & 850 & 0.21 & 1.2--2700 & 77 & 26, 97\tablenotemark{c} & {\bf 0.23} & {\bf 0.84} & {\bf 0.9 $\pm$ 0.1} & {\bf 2.0 $\pm$ 0.3} & 1 \\
NGC~7538 & 450 & 0.11 & 4--3100 & 67 & 31, 120\tablenotemark{c} & {\bf 0.31} & {\bf 0.92} & {\bf 0.9 $\pm$ 0.1} & {\bf 2.6 $\pm$ 0.8} & 1 \\
$\rho$Oph & 1300 & 0.0085 & 0.05--3.2 & 58 & 0.5 & 0.5 & 1.5 & {\bf1.5} & {\bf2.5} & 2\\ 
$\rho$Oph & 850 & $\sim$0.025 & 0.02--6.3 & 55 &0.6 & {\bf 0.5} & {\bf 1.0-1.5} & 1.5 & 2.0-2.5 & 3\\
NGC 2068/2071 & 850 & 0.02 &$\sim$ 0.3--5 & 70 & 0.7 & {\bf 0.5} & {\bf 1.1} & 1.5 & 2.1 & 4 \\
Orion B\tablenotemark{d} & 850 & 0.03 & 0.06--30.3 & 75 & $\sim 1.0$ & {\bf 
0.5} & {\bf 1.5} & 1.5 & 2.5 & 5 \\ \tableline \multicolumn{10}{c}{Single Power Law Fits} \\ \tableline
Serpens & 3000 & $\sim$0.0075 & $\sim$ 0.4--32 & 26 & \nodata & \nodata & 1.1 & \nodata & {\bf 2.1} & 6 \\
Lagoon (M8) & 450 \& 850 & 0.13 & 6.3--43 & $\sim$ 50 & \nodata & \nodata & {\bf 0.7 $\pm$ 0.6} & \nodata & 1.7 $\pm$ 0.6 & 7 \\
KR 140 & 850 & 0.22 & 0.7--130 & 25 & \nodata & \nodata & {\bf 0.49 $\pm$ 0.04} & \nodata & 1.49 $\pm$ 0.04 & 8 \\
RCW 106 & 1200 & 0.42 & $\sim$ 40--10$^{4}$ & 95 & \nodata & \nodata & 0.6 $\pm$ 0.3 & \nodata & {\bf 1.6 $\pm$ 0.3} & 9 \\
\enddata
\tablenotetext{a}{Number of clumps included in the fit.}
\tablenotetext{b}{Cumulative and differential mass function exponents.  Values calculated by 
the respective authors are in bold.  The other values were calculated by us, assuming $\gamma = \alpha+1$.}
\tablenotetext{c}{Break point masses determined from the cumulative and differential mass functions, respectively.}
\tablenotetext{d}{Includes NGC 2068, NGC 2071, and HH 24/25/26.}
\tablerefs{(1) This paper; (2) \citet{man98}; (3) \citet{dj2000b}; (4) 
\citet{m01}; (5) \citet{dj2001}; (6) \citet{ts98}; (7) \citet{tot}; (8) \citet{kerton}; (9) \citet{mookerjea}}
\end{deluxetable}

\end{document}